\documentclass[sigconf, bookmarksnumbered,unicode,dvipsnames]{acmart}

\usepackage{booktabs}
\usepackage{subfigure}
\usepackage[utf8]{inputenc}
\usepackage{hyperref}
\usepackage{graphicx}
\usepackage{epsfig}
\usepackage{color}
\usepackage{comment}
\usepackage{float}
\usepackage{booktabs}
\usepackage{caption}
\usepackage{array}
\usepackage{color}
\usepackage{xcolor}
\usepackage{url}
\hypersetup{
	colorlinks,%
	citecolor=blue,
	filecolor=black,%
	linkcolor=blue,%
	urlcolor=blue}
\usepackage[english]{babel}
\usepackage{multirow}
\usepackage{balance}

\usepackage{amsmath}
\usepackage{amsfonts}
\usepackage{amssymb}
\usepackage{bm}
\usepackage[flushleft]{threeparttable}

\usepackage[draft]{todonotes} 

\usepackage[misc,geometry]{ifsym} 

\newcommand{\rowgroup}[1]{\hspace{-0.7em}#1}

\newcommand{\oHTTP}{unpatched HTTP server}

\newcommand{\iSSL}{insecure SSL implementation}
\newcommand{\HOC}{\texttt{HttpOnly} cookie}
\newcommand{\XFO}{\texttt{X-Frame-Options}}

\newcommand{\XCTO}{\texttt{X-Content-Type-Options}}
\newcommand{\MCI}{Mixed-content inclusions}
\newcommand{\mCI}{mixed-content inclusions}
\newcommand{\SC}{\texttt{Secure} cookie}
\newcommand{\CSP}{\texttt{Content-Security-Policy}}
\newcommand{\HSTS}{HTTP \texttt{Strict-Transport-Security}}
\newcommand{\SSVF}{SSL-stripping vulnerable form}
\newcommand{\WBXP}{Weak browser \texttt{XSS} protection}
\newcommand{\wBXP}{weak browser \texttt{XSS} protection}

\begin{document}

\fancyhead{}
\settopmatter{printacmref=false, printfolios=false}


\title{Herding Vulnerable Cats: A Statistical Approach to Disentangle Joint Responsibility for Web Security in Shared Hosting}


\author{Samaneh Tajalizadehkhoob
\href{mailto:s_DOT_t_DOT_tajalizadehkhoob_AT_tudelft_DOT_nl}{\Letter}}
\affiliation{\institution{Delft University of Technology}}

\author{Tom van Goethem}
\affiliation{ \institution{imec-DistriNet, KU Leuven}}

\author{Maciej Korczy\'nski}
\affiliation{\institution{Delft University of Technology}}

\author{Arman Noroozian}
\affiliation{\institution{Delft University of Technology}}

\author{Rainer B\"{o}hme}
\affiliation{\institution{Innsbruck University}}

\author{Tyler Moore}
\affiliation{\institution{The University of Tulsa}}

\author{Wouter  Joosen}
\affiliation{\institution{imec-DistriNet, KU Leuven}}

\author{Michel van Eeten}
\affiliation{\institution{Delft University of Technology}}


\begin{abstract}

Hosting providers play a key role in fighting web compromise, but their ability to prevent abuse is constrained by the security practices of their own customers. {\em Shared} hosting, 
offers a unique perspective since customers operate under restricted privileges and providers retain more control over configurations. We present the first empirical analysis of the distribution of web security features and software patching practices in shared hosting providers, the influence of providers on these security practices, and their impact on web compromise rates. We construct provider-level features on the global market for shared hosting -- containing 1,259 providers -- by gathering indicators from 442,684 domains. Exploratory factor analysis of 15 indicators identifies four main latent factors that capture security efforts: content security,  webmaster security, web infrastructure security and web application security.  We confirm, via a fixed-effect regression model, that providers exert significant influence over the latter two factors, which are both related to the software stack in their hosting environment. Finally, by means of GLM regression analysis of these factors on phishing and malware abuse, we show that the four security and software patching factors explain between 10\% and 19\% of the variance in abuse at providers, after controlling for size.
For web-application security for instance,  
we found that when a provider moves from the bottom 10\% to the best-performing 10\%, it would experience 4 times fewer phishing incidents.
We show that providers have influence over patch levels--even higher in the stack, where CMSes can run as client-side software--and that this influence is tied to a substantial reduction in abuse levels.

\end{abstract}





\maketitle

\section{Introduction}

Global web infrastructure is compromised at scale in support of a myriad of cybercrime business models, from phishing to botnet command and control (C\&C) to malware distribution. The responsibility for remediating compromised resources is shared between webmasters and multiple infrastructure operators, notably hosting providers, domain name registrars and internet service providers (ISPs). The important role of hosting providers is codified in best practices from industry organizations such as M3AAWG and SANS \cite{SANS-bestprac,hosting-bestprac,M3AAWG}. These guidelines encourage providers to take sensible steps, such as keeping customer software updated.

When the defenses fall short and resources are compromised, providers are regularly faulted for not doing enough to forestall compromise (e.g., \cite{canali2013role,stone2009fire}).  
This raises the question, however, of what providers can realistically achieve in terms of preventing abuse. Compromise rates are driven by many factors outside the immediate control of providers, not least of which is the security decisions and patching practices of their own clients~\cite{Li16,li2016you}. 
It is this joint responsibility between providers and webmasters that makes answering the question so difficult. In this paper, we provide an answer for the case of {\em shared} hosting, one of the most prevalent and affordable ways to publish web content in which many websites share the same server. 

We focus on shared hosting services for several reasons. First, its customers operate under restricted privileges. Hosting providers maintain administrator privileges and can typically regulate what software is installed and whether it is updated. As acknowledged in M3AAWG's best practices, providers have the most control over, and hence most responsibility for, their resources in shared hosting plans, compared to other hosting services ~\cite{M3AAWG}. Second, even when customers can change configurations, shared hosting providers maintain a strong influence by provisioning default configurations that may or may not be the secure. 

Put another way, if hosting providers can and do make a difference in improving security, we would expect to find evidence for it in this segment of the market. 
Third, this segment matters in the overall scheme of web compromise. Shared hosting is associated with especially high concentrations of abuse~\cite{apwg2014,apwg2015,nomshosting2016}. In the data examined for this paper, for example, around 30\% of all abused domains were on shared hosting.

Another barrier to assessing provider efforts to prevent abuse is that their efforts cannot be measured directly. We cannot, for example, measure each provider's security budget, abuse team staff levels, or uptake of technologies to mitigate attacks. In economics terms, there is an inherent information asymmetry about the extent and efficacy of the security efforts undertaken by providers. 

We overcome this barrier by adopting a new approach, adapted from psychometrics, that constructs an indirect measure of security effort by amalgamating a disparate set of observable features such as patching levels and secure web design practices. There are two key benefits of our approach. First, we do not presume ex ante if it is the webmaster or hosting provider who is responsible for these features. Who drives patching of Content Management Systems (CMSes), for example? Rather than make a priori assumptions, we answer these questions empirically and thereby deal with the joint responsibility problem. Second, we do not presume a direct causal relationship between the observable features and how the website is ultimately compromised. For example, setting a Content Security Policy may not stop compromise, yet its presence does reflect the security efforts put into increasing website defences.

\medskip \noindent  We make the following contributions:
 
\begin{itemize}
\item We present the first comprehensive measurement study of the population of shared hosting providers, revealing patterns in 15 indicators spanning domain security and software patching efforts, captured from a sample of 442,684 domains across 1,259 providers.
\item We find that most discovered installations of webservers and admin panels (87\%) and (70\%) were running unpatched versions. In stark contrast, CMS installations were unpatched in just 35\% of cases. This perhaps reflects a difference in the probability of compromise between lower and higher levels of the software stack. Version hiding is a widespread hardening effort--e.g., 66\% of admin panel installations hide version information. By contrast, indicators of domain security, such as \HOC~ and \texttt{Content-Security-Policy}, are rare (13\% and 0.2\% of domains, respectively).
\item We demonstrate a new statistical approach to empirically disentangle the contributions of different parties to a joint security outcome. Different from prior research, we do not make ex ante assumptions about the meaning of security indicators (e.g., that their configuration is under the control of the providers and accurately reflect their efforts). Instead, we use the indicators to induce latent factors that can be interpreted and empirically attributed to roles of responsibility. We then regress these factors on measurements of compromise, while controlling for exposure. 
This approach can be adopted to study other areas of joint responsibility, such as between cloud hosting providers and tenants, or corporate system administrators and end users.

\item 
We find that webmaster and web application security efforts significantly reduce phishing and malware abuse. For example, the best-performing 10\% of providers (in terms of web application security effort) experience 4 times fewer phishing incidents than the bottom 10\% of providers. Moreover, we find that providers can influence patching levels, even for software running at the application level such as CMSes. The providers that do a better job of patching their customers see reduced rates of compromise. This provides the first compelling statistical evidence of the security benefits of hosting providers adhering to industry best practices.
\end{itemize}

The paper proceeds as follows:
Section \ref{sec:data} explains the data and methodology used to sample domains, identify shared hosting providers, estimate their size, and measure compromise rates. Section \ref{sec:features} outlines the details of our active measurement setup and describes the effort-related features we collected.
Section \ref{sec:desc} presents an empirical view of the web security landscape in shared hosting.
Section \ref{sec:dead} discusses the reasoning behind
why the collected features should not be used as direct causal explanations of abuse, highlighting the need for latent variables.
Section \ref{sec:factor_analysis} explains the statistical approach to estimate the latent variables and to empirically disentangle the contributions of different parties to a joint security outcome.
Section \ref{sec:model}, we assess the impact of the latent factors on abuse incidents.
Section \ref{sec:limitations} discussed the limitations of our study and section~\ref{sec:related_work} revisits related work. 
Finally, we discusses our main conclusions and implications in Section \ref{sec:conclusions}.

\section{Data}
\label{sec:data}

\paragraph{Shared hosting providers}
Our first task is to identify the population of shared hosting providers.
We build on the procedure developed in prior work~\cite{nomshosting2016,toit}. 
We start by populating a list of all domain names\footnote{We define domain name as a second-level or third-level domain, depending on whether the relevant TLD registry provides such registrations, e.g., \texttt{example.pl}, \texttt{example.com.pl}, 
\texttt{example.gov.pl}, etc.} 
and their IP addresses that were observed by DNSDB -- a large passive DNS database\footnote{\url{https://www.dnsdb.info}} -- in March 2016. Subsequently, we mapped the IP addresses to the organizations to which they are allocated using MaxMind's WHOIS API~\footnote{\url{http://dev.maxmind.com}}. From the resulting list of organizations, we group very similar provider names as single companies. Next, we applied a series of keyword filters to weed out non-hosting services, such as DDoS-protection services, internet service providers, and mobile service providers, as discussed in~\cite{nomshosting2016}.
This provides us with a set of hosting providers. 
Next, we mark a provider as a {\em shared} hosting provider if we observe at least one IP address that hosts more than 10 domains. 
We adopt the same threshold used in other studies~\cite{vasekriskfactor,nomshosting2016}. Using an elbow plot of domain density per IP address, we confirmed a sharp increase in  density beyond a threshold of 10 to 15 domains per IP address.
The result is a global list of 1,350 shared hosting providers. 

\paragraph{Domain sample}
From the total set of 110,710,388 domains on shared hosting, we randomly sampled 500 domain names for each provider. We scanned them to verify these were still operational\footnote{Domains are sampled only from IPs marked as shared, since a provider can have shared servers next to dedicated ones}.
If fewer than 100 domains were up and running, the provider was excluded from the list (91 providers were excluded).
It should be noted that before drawing the random selection of domains, we dismissed around 4,000 parked domains, following the detection methodology outlined in \cite{vissers2015parking}.
This is specifically because a majority of  parked domains are very similar to each other (share the similar content) and typically a single webmaster owns numerous parked domains, as indicated by Vissers et al.~\cite{vissers2015parking}.
Therefore, if taken into account, the analysis is more likely to be biased towards a handful of website administrators owning a large number of domains.
By excluding parked domains, we maintain an unbiased observation of the features that are related to the efforts of the webmaster.
Accordingly, our final set contains 442,684 domains distributed over 1,259 hosting providers, located in 82 countries all over the world.

\paragraph{Size of hosting providers}
\label{ssec:sohp}
Shared hosting providers differ vastly in size, a fact to be controlled for when analyzing abuse with providers as units of analysis. Clearly, a million-site business is likely to observe more abuse than one with a few thousand customers. Unfortunately, there is no authoritative source for provider size. To estimate it from the available data, we use two different size indicators, each capturing a different aspect of the shared hosting providers. {\em Shared hosting IP space size} is the number of IP addresses hosting at least 10 or more domains. It is calculated by summing up all the IP addresses defined as shared, associated with domain names per provider that have been observed in the 
passive DNS data. 
The mean, median and maximum values are 636, 137 and $71,448$ respectively, across  providers in our sample.
{\em Shared hosting domain space size} is the number of domains hosted on shared IPs by a particular provider. It is calculated as the sum of the domains that are associated with shared IP addresses of the provider, as seen in the DNSDB data. 
The mean, median and maximum values are  94,118, 10,233 and $3.3*10^7$ respectively, across  providers in our sample.
Note that due to a large variance and skewed distribution of the size variables, 
a log-transformation of these variables (base 10) is used in the  regression analyses of Section \ref{sec:model}.

\paragraph{Abuse data}

To estimate the compromise rate for each shared hosting provider, we used two abuse datasets. We extracted all entries that were associated with the shared hosting IP addresses of the providers and counted the number of unique domains per provider.

The {\bf phishing} data is collected from two sources: the Anti-Phishing Working Group (APWG)~\footnote{\url{http://www.antiphishing.org}} and Phishtank~\footnote{\url{https://www.phishtank.com}}. 
Both datasets contain IP addresses, fully qualified domains, 
URLs of phishing pages, blacklisting times, and additional meta-data.
For the second half of 2016, the data consisted of 62,499 distinct domains, which resolved to 47,324 IP addresses at the time of reporting.
49,065 of these domains were hosted by one of 968
shared providers in our study (The remaining 291 providers did not record any phishing during the period.)

We include drive-by-download {\bf malware} URLs flagged by the Google Safe Browsing program, as reported to StopBadware\footnote{\url{https://www.stopbadware.org}}. For the second half of 2016, there were 362,069 distinct domains newly flagged with malware. Of these, 332,625 resolved to an IP address at the time of reporting. The rest was likely maliciously registered and already suspended. Of all resolvable domains, 97,872 were hosted by one of 1,050 
shared providers in our study (The remaining 209 providers did not record any malware during the period.)
The high proportion in both datasets underscores the importance of shared hosting in distributing web-based phishing and malware.

\section{Measurement of Features}
\label{sec:features}

\subsection{Measurement setup}
\label{subsec:messetup}

We aim to collect a wide range of features, composed of vulnerabilities and weaknesses, security mechanisms, 
and software patching practices, all of which can help us estimate the amount of effort going into securing domains.

We perform a large-scale measurement to obtain information from the 442,684 sampled domains.
More precisely, we instructed our crawler, which is based on the headless browser PhantomJS\footnote{\url{http://phantomjs.org/}}, to visit up to 20 web page for each domain.
The list of web pages for a certain domain were obtained by following links starting from the home page, until either the maximum number of page visits was reached, or no further links could be discovered.

In order to restrict the feature collection process to the target domains, the crawler only considered web pages with the same second-level domain name. If, for example, the target domain \texttt{example.com} immediately redirects users to \texttt{website.com}, only a limited set of features could be obtained, i.e., server-level features and those based on response headers sent out by \texttt{example.com}. This was done to ensure that only information related to the website hosted in the shared hosting environment was considered.
In total, it took our crawler, which was distributed over 15 virtual machines, each composed of 4 CPUs and 4GB RAM, 7 days to visit and extract information from the 7,463,682 web pages.

We gather information to construct a list of 15 features, which is an extension of the web-based security features explored in prior work~\cite{van2014large}. 
Our features give an indication of both security-related configurations, such as the deployment of \texttt{Content-Security-Pol\\icy},
 and patching practices of various software such as CMSes, admin panels, PHP and SSH.
Consequently, the captured features reflect security practices employed by both the shared hosting providers as well as the domain owners (webmasters) themselves. 
In the following sections, we briefly discuss these two groups.
For the extensive list of features, please refer to Table~\ref{tab:summary_features}.

Note that for most of the collected features, we do not expect to observe a direct causal relation on abuse practices. Instead, we consider the features to be proxies of the efforts made by the providers and webmasters. We discuss the limitations of treating these features as direct indicators of effort in greater detail in Section~\ref{sec:dead}.

\noindent \paragraph*{Ethical considerations} 
We designed our measurement techniques to be as unobtrusive as possible. We collected no more data than necessary and carefully scheduled our requests so that no single server could be overloaded. All features were obtained through passive observation and we added various countermeasures to prevent any irregular interactions with third party websites. Finally, we report the findings in an anonymized manner.


\subsection{Domain security indicators}
\label{subsec:secfeatures}

As domains are prone to a large variety of potential vulnerabilities and weaknesses, the web security community has for a long time supported hosting providers and webmasters with mechanisms that enable them to apply a defense-in-depth approach. In this section, we discuss how we collect a multitude of security-related features to get an approximation of security efforts for domains.

Cross-site scripting (XSS) vulnerabilities are among the most critical security risks according to OWASP~\cite{OWASP,weichselbaum2016csp}.
We look for the presence of the \texttt{Content-Security-Policy} response header, as it can be used to protect against XSS attacks.
We consider a domain to have \wBXP~if an administrator has disabled the default browser mechanism to detect and block reflected XSS attacks by means of the \texttt{X-XSS-Protection} response header.
We also check for the presence of \texttt{HttpOnly}, which helps reduce the potential consequences of XSS attacks, and \texttt{X-Frame-Options}, which can be used to thwart clickjacking.
In addition, we check if the \texttt{Secure} cookie attribute and the {\HSTS} response header are present, as they both can effectively improve transport layer security.
Properly implemented web applications are also crucial. We define the \SSVF~ feature when a website has a form (e.g.\ on a login page) pointing to an HTTPS endpoint while being loaded over an insecure connection. Accordingly, the \mCI~happen when a website's content (e.g.\ JavaScript code, style-sheets, images etc.) is included over an insecure connection, while the web page was loaded over SSL/TLS.

Note that we indicate the direction of the features by ($-$) and (+) signs in Table~\ref{tab:summary_features} since not all features have a positive effect, such as \mCI{}, \SSVF{}, and \wBXP.

\subsection{Software patching practices}
\label{subsec:softfeatures}

In addition to the security mechanisms discussed in the previous section, the act of patching server software and web applications plays a crucial role in the security posture of websites.

Often, attackers exploit known vulnerabilities present in unpatched software (e.g., vulnerabilities reported in the National Vulnerability Database~\cite{NVD}).
Therefore,  it is generally considered best practice for providers as well as webmasters to employ patch management mechanisms regularly and extensively.

Content Management Systems (CMSes) have been amongst the most exploited software stacks for many years~\cite{vasekriskfactor,soska2014automatically,IBMcmshijacking}.
Depending on the administration rights in the shared hosting environment, CMSes can be updated either by the webmaster or the shared hosting provider herself.
In this paper, we limit our scope to the CMSes with the majority of market share, namely WordPress, Joomla!\ and Drupal CMSes~\cite{cmsmarket}. 

The presence and version number of these three CMSes are determined in two phases: first, a basic scan is performed using our crawler which tried to infer the version number from the \texttt{<meta name=$''$generator$''$>} HTML tag.
However, as many CMSes allow hiding the version number, something that is generally considered a good practice against automated attack scripts, we perform a second, more comprehensive scan.
For the comprehensive scan we made use of well-known industry tools such as Sucuri WPScan\footnote{\url{https://wpscan.org}} and WhatWeb\footnote{\url{https://whatweb.net}}.
For the latter, we updated the original scripts to allow us to incorporate the latest versions of the targeted CMSes.

In addition to the experiment that determines the presence and version number of CMSes, we performed a similar experiment that focused on admin panels, a type of technology that is innate to the shared hosting environment.
In this paper we focus on the four most popular admin panels, namely cPanel, Plesk, DirectAdmin, and  Virtualmin.
We instructed our crawlers to visit the domains at the port numbers that are typically associated with the admin panels, e.g.\ port 2082 and 2083 for cPanel.
We then improved our measurements by visiting the endpoints that we found to often be used as a shorthand to link to the admin panel, e.g.\ \texttt{/panel/}.
Based on the response headers, HTML contents, and redirection chains that were captured by our crawlers, we tried to determine the presence and, when possible, the version number of the admin panels.
This allowed us to obtain the version information for approximately 33\% of the domains with admin panel in our sample.

Moreover, other components that contribute to the software stack, such as the HTTP server, SSH server and PHP interface, should also be treated as part of the threat surface.
In this paper we focus on Apache, Microsoft IIS and nginx for the HTTP servers.
For the features related to the infrastructure of the web host, 
we inferred the version information through either the \texttt{Server} and \texttt{X-Powered-By} response headers (for webserver and PHP), or by the banner that was returned, e.g.\ the banner on port 22 for SSH.

Lastly, we look into SSL/TLS implementations, as they are important to prevent attacks on the transport layer. 
To assess weaknesses in the SSL/TLS infrastructure, we used sslyze~\cite{sslyze}.
The domain's SSL/TLS implementation is considered insecure when it was vulnerable to Heartbleed, supports old protocols (SSLv2, SSLv3), enables compression, or is vulnerable to CCS injection~\cite{ccsinjection}.

For all software where the  version number could be determined by our scanner, we make the distinction between software that is patched and unpatched.
Generally, we consider a software version to be \emph{patched} if it was packaged in one of the supported versions of OSes with larger market share namely, Ubuntu, Debian, and CentOS~\cite{OSmarketshare} at the time of our the measurement (November 2016).
This approach is relatively generous in considering software patched: patches are often backported to older versions; as we did not undertake any intrusive actions to determine the patch-level of software, no distinction is made between versions with or without these backported patches.
Note that all the older versions of software packaged in OSes are deprecated and contain vulnerabilities. For instance, PHP version 5.3.2 had a vulnerability (CVE-2012-2317) that would allow attackers to bypass authentication in applications that would otherwise be secure. This was then patched in the later versions packaged. A more recent example is CVE-2015-8867 in certain versions of PHP (5.4.44, 5.5.x before 5.5.28, and 5.6.x before 5.6.12)~\cite{NVD}.
A list of software and their patched versions is included in the Appendix.

Due to the automated nature of our experimental setup, the measurements may be subject to certain limitations.
Despite the preventive measures we have taken to make the generated web traffic reflect the browsing behavior of a regular user, there could still be providers who will block our scanning attempts.
Moreover, it is possible that certain software was not found within the scanning threshold due to hardening techniques. More specifically for admin panels, if the software was not located at a default location, we would not be able to detect it.
Furthermore, as we focus on a limited set of software, it is possible that a domain makes use of a different software stack, or that it was hand-constructed.

\begin{table}[!ht]
\caption{Summary of measured domain security and software patching indicators in absolute and relative terms.}
\label{tab:summary_features}
\setlength{\tabcolsep}{12pt}
\resizebox{\columnwidth}{!}{%
\begin{tabular}{lrr}
\toprule
\rowgroup{Feature} & \rowgroup{\# of domains}  &\rowgroup{\% of domains}   \\
\midrule
\rowgroup{HTTP server}& \rowgroup{398,929} & \rowgroup{90.11} \\
no version information & 195,474 & 44.15 \\
Patched version & 58,818 & 13.28 \\
Unpatched versions & 144,637 & 32.67 \\[0.4em]

\rowgroup{SSL}& \rowgroup{288,018} & \rowgroup{65.06}\\ 
Patched version & 206,680 &  46.68 \\
Unpatched versions & 81,338 &  18.37 \\[0.4em]

\rowgroup{Admin panel}& \rowgroup{178,056} &  \rowgroup{40.22} \\
no version information & 118,768 &  26.82 \\
Patched version &17,949&  4.05 \\
Unpatched versions &41,600&  9.39 \\[0.4em]

\rowgroup{PHP}& \rowgroup{156,756} & \rowgroup{35.41} \\
Patched version & 47,596 & 10.75 \\
Unpatched versions & 109,160 & 24.65 \\[0.4em]

\rowgroup{OpenSSH}& \rowgroup{130,146} & \rowgroup{29.39} \\
no version information & 716 & 0.16 \\
Patched version & 36,444 & 8.23 \\
Unpatched versions & 92,986 & 21.00 \\[0.4em]

\rowgroup{CMS} & \rowgroup{103,741} &\rowgroup{23.43} \\
no version information & 10,043 & 2.26  \\
Patched version &    61,457 & 13.88  \\
Unpatched versions &    32,264 & 7.28 \\[0.4em]

\rowgroup{\HOC} (+) &   \rowgroup{57,696} &   \rowgroup{13.04} \\
\rowgroup{\XFO} (+) &     \rowgroup{22,212} &  \rowgroup{5.02}\\
\rowgroup{\XCTO} (+) &    \rowgroup{8,685} &  \rowgroup{1.96} \\
\rowgroup{\MCI} ($-$) &   \rowgroup{2,107} &   \rowgroup{0.47} \\
\rowgroup{\SC} (+) &       \rowgroup{1,378} &     \rowgroup{0.31} \\
\rowgroup{\CSP} (+) &     \rowgroup{894} &     \rowgroup{0.20} \\
\rowgroup{\HSTS} (+) &    \rowgroup{847} &     \rowgroup{0.19}\\
\rowgroup{\SSVF} ($-$) &  \rowgroup{515} &     \rowgroup{0.11}\\
\rowgroup{\WBXP} ($-$) &  \rowgroup{376} &    \rowgroup{0.08} \\

\midrule
\end{tabular}
}
\end{table}

%

\section{Descriptive Findings about the Landscape}
\label{sec:desc}

Previous research has explored individual security features at the domain level. 
We now extend this approach in two ways: by combining these features with software patching practices and by moving from individual domains to the level of providers.
What is the prevalence of security features across domains and providers? How patched are software installations? Do patching rates vary substantially from one provider to the next? Do different portions of the software stack have different updating behavior?


\subsection{Distribution of security features}
Table~\ref{tab:summary_features} presents a summary of the distribution of all security features, both positive and negative. 
The security features are presented as boolean variables, with 1 pointing to the direction of the variable.
The first column indicates the total number of domains with a particular feature and the second column reports the percentage of all domains with this feature. 

The overall pattern is clear. Across the landscape, although crucial, the positive security indicators have low to almost negligible adoption rates. Out of 442,684 scanned domains, {\HOC} reaches a somewhat respectable 13\%, but after that the prevalence drops quickly. Two features are present in less than 0.3\% of all domains. The good news is that the observed negative security features that can result in vulnerabilities are equally sparse: {\MCI} is the most widespread at 0.5\%. 

\begin{figure}[ht!]
\centering
\includegraphics[width=0.5\textwidth]{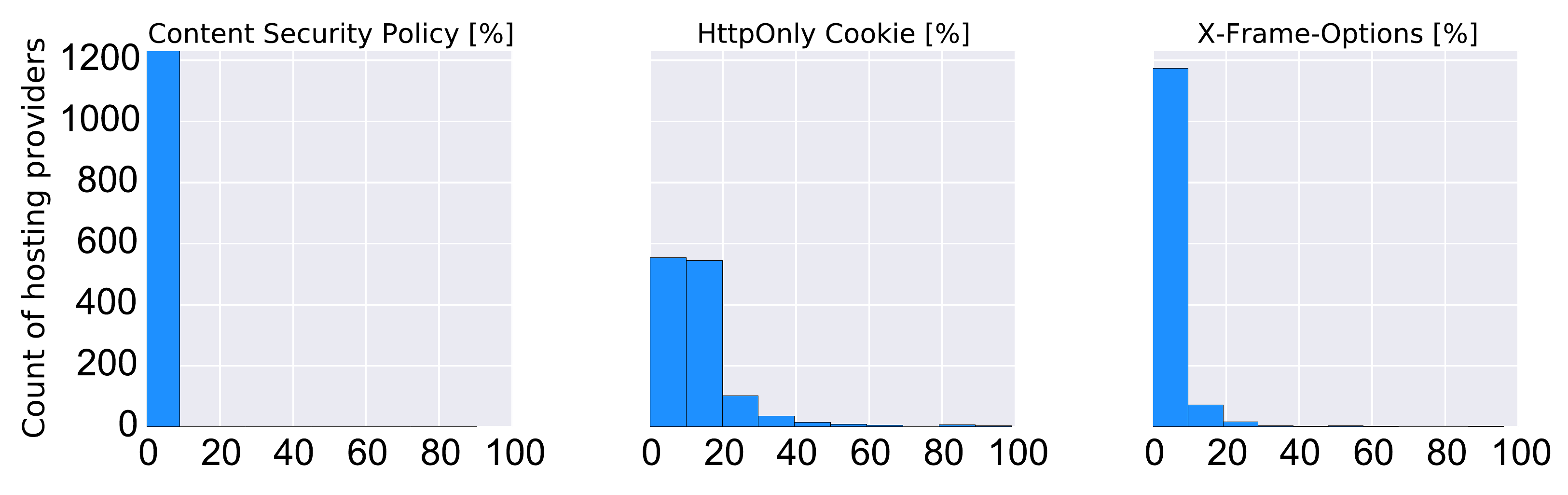}
  \caption{Distribution of security features over hosting providers}
  \label{fig:somefeatures_org_dist}
\end{figure}

To illustrate, Figure~\ref{fig:somefeatures_org_dist} displays the percentage of domains at a provider that have {\CSP}, {\HOC} or {\XFO}.
At most providers, only a small fraction of their domains support these features, hence one sees rarely any large concentration of a feature within a group of providers.
In fact, for 1,100 providers (95\% of the providers we evaluated), fewer than 20\% of their domains in our sample have {\HOC} enabled. 
The exception is a group of 9 providers where 80\% of the domains have {\HOC} enabled, indicating a provider effort in the form of provisioning default secure configurations.
To further validate this assumption, we tried to contact this set of nine providers manually and check whether they provide certain security features by default. We have been contacted back by three of the providers. Two of the providers confirmed that depending on the customers, they might set {\HOC} by default in the cases where they are the responsible entity for the customer's security.
Another provider pointed out that the default {\HOC} setting is a built-in feature in the \texttt{DotNetNuke} CMS they employ.

Note that the median and mean complexity of the webpages in our sample (measured by the number of endpoints)  are 11 and  71.68, respectively. Having that in mind, we expect some of these features are only useful in specific configurations, so widespread adoption is not to be expected. 
Not every page will set a cookie, for example, and not every cookie needs to have the \texttt{Secure} or \texttt{HttpOnly} attribute. A cookie might set a language preference, it might need to be accessible in JavaScript, and it does not matter if this leaks in a man-in-the-middle  attack. 
Also, for {\XFO}, it makes sense that this header is only added on pages that are subject to clickjacking attacks.
On the other hand, features such as {\CSP} would benefit many domains and, as other work has noted~\cite{calzavara2016content}, adoption is disappointingly low. 

Of all providers, only 6\% has more than a single domain with {\CSP} in the sample. 
That being said, there is an interesting long tail for these scarce features, where the provider seem to play a role. For instance, the managed hosting provider \texttt{Netsolus.com}, has more than 92\% of its domains in our sample enabled with \CSP~ and \HOC~, which again suggests a provider wide setting rather than effort of individual webmasters.

\subsection{Distribution of software patching features}
\begin{figure}[!htp]
\centering
\includegraphics[width=0.35\textwidth]{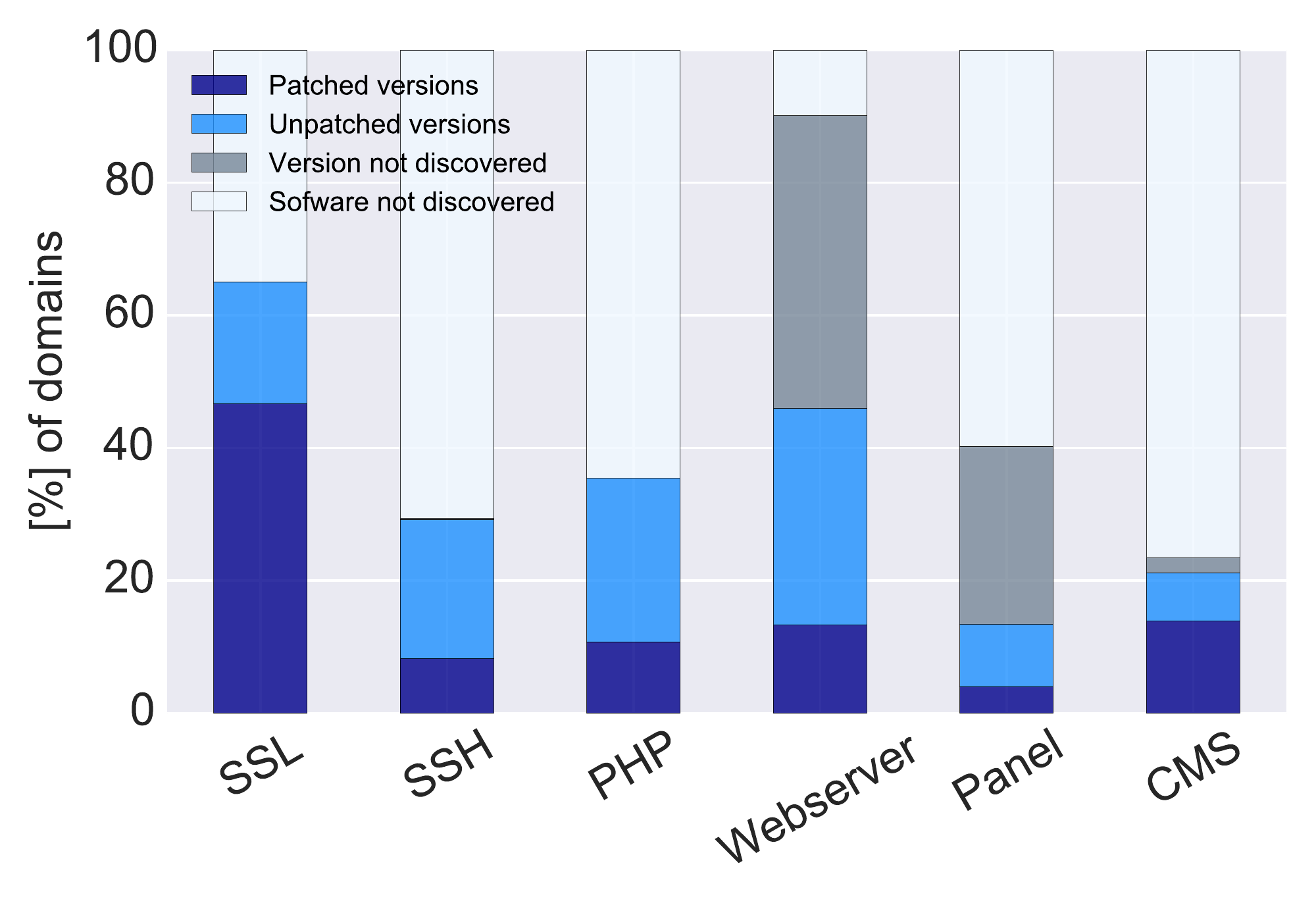}
  \caption{Software patching distribution across domains}
  \label{fig:software_outdated}
\end{figure}

Regarding software installations, Figure~\ref{fig:software_outdated} provides a visual overview of the data in Table~\ref{tab:summary_features}.
The colored area shows the portion of all domains where we were able to discover a certain type of software. This is subdivided in installations where we found the patched version (dark blue), where we found an unpatched version (light blue) and where we could not identify the version (grey).

Manual analysis of the software patching features reveals several interesting patterns.
In the rest of this section, we discuss software discoverability by attackers and version hiding efforts by defenders. Then we look at the state of patching across the web stack.

\subsubsection{Hardening practices}
Discovering the presence and version of a software installation on a domain is more than a measurement problem. The techniques we deploy can also be used by attackers seeking vulnerable targets, especially if they scale easily. This incentivizes defenders to harden software installations to be less trivially discoverable and to not reveal version information.

\begin{figure}[ht!]
\centering
\includegraphics[width=0.30\textwidth]{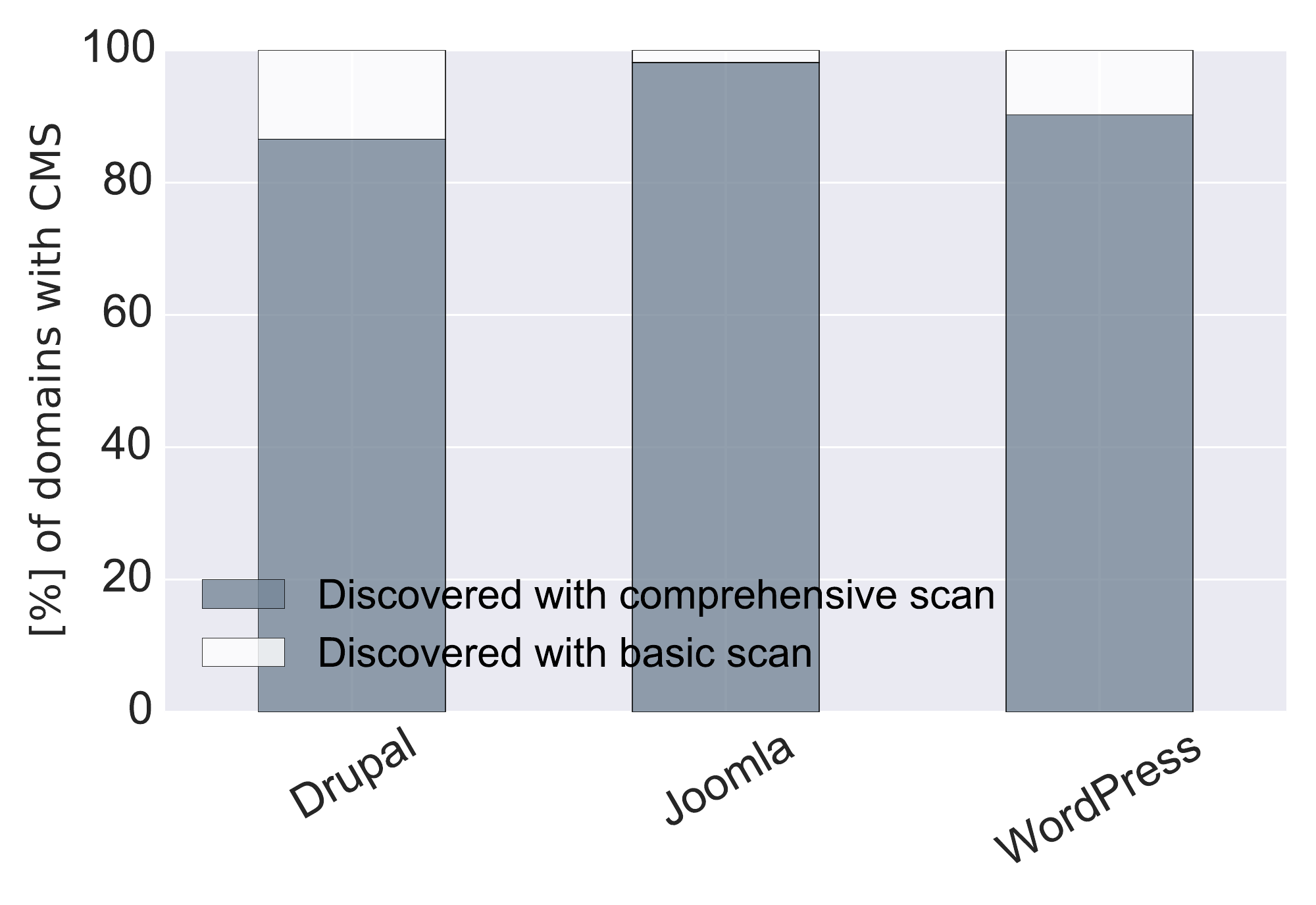}
  \caption{Portion of CMS installations discovered via basic vs. comprehensive scans}
  \label{fig:software_scantype}
\end{figure}

Indeed, in the case of the three main CMSes, a basic scan was rarely effective. Figure~\ref{fig:software_scantype} shows that most installations were discovered only through more intrusive industry tools, described in Section~\ref{subsec:softfeatures}.
Overall, 23\% of the domains had one of the three main CMSes installed. 
To determine the validity of our results, we manually inspected 40 domains per CMS type, both from domains for which we discovered an installation and from those for which we did not. We found one false positive where the domain did not have any CMS and no false negatives. Most of the pages that we marked as no CMS pages were either static HTML or featured some custom-made CMS.
It is an open question as to what made the discovery more difficult: webmaster action, provider action, or the default configuration provided by the software vendor.

Similarly to CMS, most of the well-known admin panels were only discovered after a  more comprehensive scans. We found them on 40\% of the domains. In a shared hosting environment, admin panels seem a necessity, so the actual prevalence is likely to be higher. Many providers, however, appear to shield them from being discovered, even by more comprehensive scans. They are using custom solutions or hide them behind customer portals with restricted access. 

Version hiding is also a popular hardening technique. 
For SSH, all version information is available, as required by the protocol.
It is interesting that PHP almost always provides version information, whereas only 50\% of HTTP webservers came with version information. Finding the version information was harder for admin panels. We managed to find it for around 32\% of all domains with one of the main admin panel packages installed.
For CMSes, version information could be obtained for around 90\% of the installations.
Given the known hardening techniques such as password-protecting the \texttt{/wp-admin/} directory, disabled PHP execution etc., we suspect that this reflects the efficacy of the industry scanning tools, rather than provider or customer practices~\cite{sucuri}.

\begin{figure}[ht!]
\includegraphics[width=0.5\textwidth]{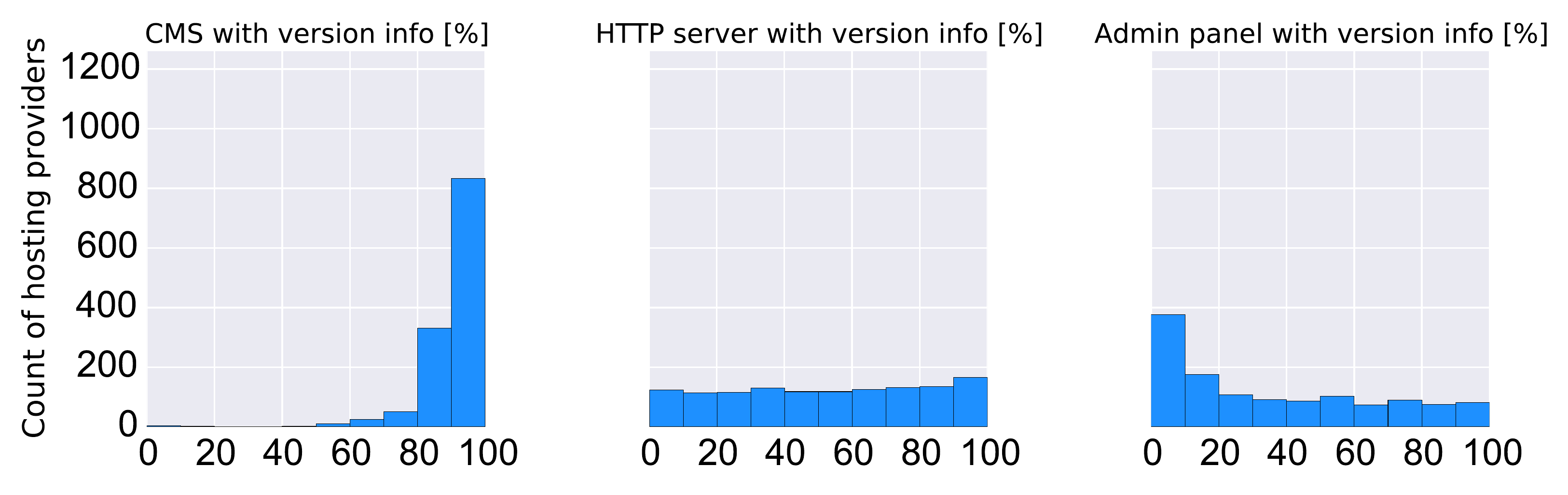}
  \caption{Distribution of discoverable software  version across providers}
  \label{fig:patching_org_distribution_withversion}
\end{figure}

We are interested in the difference among providers in version hiding efforts. 
We looked at the percentage of software installations at a provider for which version information was available. 
Figure~\ref{fig:patching_org_distribution_withversion} displays where providers are located across a range from where just 0-10\% of their installations reveal version information to where 90-100\% do. 
The resulting distributions vary considerably by software type. For CMSes, providers are clustered at the high end of the range. 
Again, this more likely reflects the efficacy of the scanning tools than of provider practices.
For webservers, however, we see a very different pattern; an almost uniform distribution across the range.
In some provider networks, nearly all versions are visible. In others, virtually none are. The rest are somewhat evenly distributed across the intermediate intervals.
If we assume that shared hosting providers have control over the web server configuration, which seems reasonable, then this distribution suggests that most providers are not consistently applying version hiding in one way or another.
The mix of both hidden and visible version information might reflect changes in the provisioning processes over time. As new servers get added, a different default setup might be in use, hiding or not hiding this information. 
For admin panels, we see yet another distribution. A concentration of providers is on the low end of the range, where version information is mostly hidden across their network. This suggests a consistent practice. But we also see a flat distribution over the rest of the range. Here, again, we might see either changing or ad hoc provisioning processes. It seems unlikely that this reflects customer action. 

\subsubsection{Patching}

More important than hiding version information is to ensure that software is not exploitable in the first place~\cite{IBMcmshijacking}.
In this section, we explore patching practices.
Figure~\ref{fig:software_outdated} displays the proportion of domains with the patched version of software, with unptached versions, and installations for which we could not determine the version.

Appendix~\ref{app:ver} lists the patched versions for each software package and its supported branches. 
We find that 19\% of domains use unpatched SSL. Note that unpatched means SSLv2 and SSLv3 or containing certain vulnerabilities such as Heartbleed, CCS injection, etc.
For PHP and SSH, it is clear that fewer domains are running the patched versions relative to the unpatched version. 
For webservers and admin panels, the majority of installations were running unpatched versions--87\% and 70\%, respectively.

In stark contrast to this stand CMS patch levels: less than 35\% were not running the latest version.
This probably reflects two interlocking mechanisms: a penalty for not updating through higher probability of compromise, as CMSes are known targets for attackers, and increasing support for auto-updating mechanisms, partly in response to these attacks.
The fact that lower layers of the software stack such as webserver and SSH do not update as aggressively suggests that the risk of compromise is lower.
This might be due to older versions still being patched internally with critical security updates or to the fact that vulnerabilities are harder to exploit remotely than in CMS software.

\begin{figure}[ht!]
\includegraphics[width=0.5\textwidth]{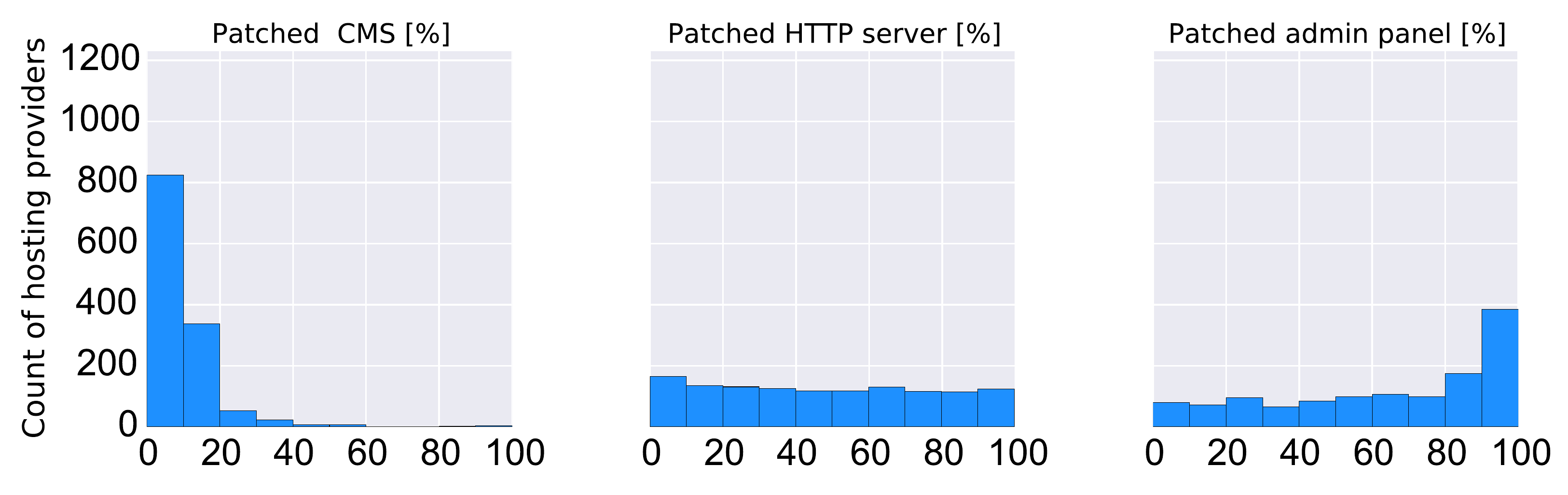}
  \caption{Percentage of domains per provider with patched software versions}
  \label{fig:patching_org_distribution}
\end{figure}

Figure~\ref{fig:patching_org_distribution} shows the proportion of domains running older versions in each provider. 
Providers are somewhat normally distributed when it comes to unpatched CMS versions in their network.
This is consistent with a natural update cycle over many different customers, each with slightly different time lags.
The distribution of providers is more uniform for web servers, which again points to changes in provisioning.
We see a positive skew for admin panels, where a significant portion of the providers have almost all installations on the latest version. 
If we assume that both webserver and admin panel software is under the provider's control, this difference is remarkable. 
It might reflect different incentives at work. Since updating incurs cost and can cause compatibility issues, providers might avoid it in the absence of a pressing need. This leaves only changes in provisioning to change the mix of software versions over time, which means the mix of latest and older versions gradually shifts, consistent with the flat distribution of webserver versions. 
For software that is attacked more often, we would indeed expect a higher concentration of providers running the patched version -- which is indeed what we see for the admin panels.

\section{Dead End: Direct Relation Between Security Indicators and Abuse}
\label{sec:dead}

Our main goal in this paper is to study the relationship between the security indicators we collected and abuse, at the level of shared hosting providers, and eventually \emph{understand} the influence of provider security effort. This justifies the choice of inductive statistical techniques which promise coefficient estimates that lend themselves to meaningful interpretation, as opposed to machine learning, which maximizes predictive power with non-linear methods. Statistical techniques produce exact (up to the arithmetic precision) solutions as well as indicators of confidence, e.\,g. in the form of significance tests. They can be calculated as a by-product of the estimation, therefore relaxing data requirements compared to heuristic cross-validation typical for machine learning.

Nevertheless, our task is complicated by the fact that each provider hosts a varying number of sites of varying functionality, complexity, exposure, and customer (i.e., webmaster) expertise. The security outcome for each site is a result of joint efforts of provider and webmaster as well as attacker behavior. On the provider level, it is the result of joint efforts of many parties. 
Therefore, it is convenient and compatible with our statistical approach to model attacker behavior as a random process, which generates counts of incidents observable in our data source.

To explain our method, we contrast it to a naive statistical approach that models the indicators as direct drivers of abuse rates. An example is displayed in Table~\ref{tab:directreg}. It reports three specifications of a count-data regression model in columns. The units of analysis are providers and the dependent variable is the number of phishing incidents in the provider's shared hosting domains.
Model~(1) is the baseline, including the two size indicators (cf.~Section~\ref{ssec:sohp}). Its Pseudo-$R^2$ value of $0.68$ highlights the importance of size control in this heterogeneous dataset.
Model~(2) tries to explain abuse with one technical indicator (of insecurity), namely the number of domains with unpatched CMS. The effect is statistically significant, and in the expected direction: the positive sign of the coefficient means that more domains (log scaled) with outdated CMS coincide with more abuse. 
However, the more comprehensive Model~(3) paints a different picture. The apparent cause of abuse is not the fact that the CMS is unpatched, but the presence of a CMS in the first place. Model~(2) missed to control for the fact that websites differ in complexity and thus risk. As a result, it detected a spurious effect in the 
``unpatched'' indicator.

\begin{table}[!htp] 
\centering 
  \caption{Quasi-Poisson GLM with Log Link Function} 
  \vspace{-0.5em}
  \label{tab:directreg} 
 \resizebox{\columnwidth}{!}{%
\begin{tabular}{@{\extracolsep{5pt}}lccc} 
\\[-1.8ex]\hline 
\hline \\[-1.8ex] 
 & \multicolumn{3}{c}{\textit{Dependent variable:}} \\ 
\cline{2-4} 
\\[-1.8ex] & \multicolumn{3}{c}{Count of phishing domains} \\ 
\\[-1.8ex] & (1) & (2) & (3)\\ 
\hline \\[-1.8ex] 
 Number of hosted domains & 1.467$^{***}$ & 1.539$^{***}$ & 1.678$^{***}$ \\ 
  & (0.083) & (0.085) & (0.078) \\ 
  & & & \\ 
 Number of IPs hosting domains & 0.690$^{***}$ & 0.672$^{***}$ & 0.472$^{***}$ \\ 
  & (0.100) & (0.100) & (0.085) \\ 
  & & & \\ 
 Number of domains with oudated CMS &  & 0.010$^{***}$ & $-$0.023$^{***}$ \\ 
  &  & (0.002) & (0.005) \\ 
  & & & \\ 
 Number of domains with CMS without version info &  &  & $-$0.019$^{***}$ \\ 
  &  &  & (0.004) \\ 
  & & & \\ 
 Number of domains with CMS &  &  & 0.015$^{***}$ \\ 
  &  &  & (0.001) \\ 
  & & & \\ 
 Constant & $-$5.596$^{***}$ & $-$6.150$^{***}$ & $-$6.743$^{***}$ \\ 
  & (0.274) & (0.314) & (0.322) \\ 
  & & & \\ 
\hline \\[-1.8ex] 
Observations & 1,259 & 1,259 & 1,259 \\ 
Dispersion   & 90    & 89  & 68  \\
Pseudo $R^2$ & 0.68 & 0.71 & 0.78 \\ 
\hline 
\hline \\[-1.8ex] 
\textit{Note:}  & \multicolumn{3}{r}{$^{*}$p$<$0.05; $^{**}$p$<$0.01; $^{***}$p$<$0.001} \\ 
 & \multicolumn{3}{r}{Standard errors in brackets} \\ 
\end{tabular} 
}
  \vspace{-1em} 
\end{table} 

These findings confirm previous work at the level of domain names \cite{vasekriskfactor}. 
The authors have concluded that \textit{i)} running popular CMS platforms (WordPress and Joomla) and \textit{ii)} running up-to-date versions of WordPress increases the odds of a domain in getting compromised. Table~\ref{tab:directreg} reflects that we find similar relationships on the provider level. In addition, we identify a statistically significant effect of hardening efforts put in place by defenders in hiding the version string. 

But does hiding version information really prevent abuse? While plausible in principle, this conclusion is too early and suffers from two issues. 
The first one is known as ecological fallacy: a relationship at the level of providers might not hold at the level of domains, i.e., the abuse may not happen at the sites where the security indicator was observed. This fallacy tells us not to interpret aggregate-level analyses as individual causal relationships. As we mainly aim to study the discretion and responsibility of providers, site-level effects need to be isolated, but not necessarily attributed to individual causal relationships. 
The second issue concerns {\it unobserved} third variables. There is a plethora of web vulnerabilities and corresponding attack vectors. Any attempt to measure them comprehensively with security indicators is futile, because each indicator may suffer from the issues demonstrated in Table~\ref{tab:directreg}. 

As a way out of this dead end, we first adopt a statistical approach common in psychometrics, where dealing with unobserved constructs has a long tradition. With this lens, hiding the version information should {\it not} be interpreted as a direct cause of less compromise, but as an indicator of {\it security effort}, a latent variable indirectly measured by many correlated indicators. The convention to use many indicators reduces the measurement error in each of them. Moreover, latent variables are implicitly defined by the composition of their indicators. The main advantage of using security effort as a latent variable is that we do not need to fully understand the causal relationship of attack and defense mechanisms throughout the global shared hosting space. Instead, it is sufficient to assume that if someone makes above-average effort to, e.g., hide version information, he also takes other
steps against attacks, which are not directly captured with indicators. This way, our results become more generalizable and robust at the same time.

In the following, we will infer from data not only one, but several latent variables measuring the security effort of different parties. This allows us to disentangle the joint responsibility using empirical data, without the need to a priori assume and impose a responsible party for each security indicator.


\section {Security Effort as a Latent Variable}
\label{sec:factor_analysis}

As argued above, constructing latent factors from the security indicators we collected is superior in terms of measuring \emph{security effort} than using the indicators on their own. 
This approach also allows us to better empirically disentangle provider vs. webmaster influence over these features.

Given the restricted administrative rights in a shared hosting environment, among the features we collected, we assume that features such as \HOC~ can be modified by webmasters as well as providers, whereas other features such as HTTP web server, are more likely to be modified only by the provider itself. 

However, this statement is speculative and is not necessarily an accurate reflection of the reality for the following reasons: 
First, as earlier work also points out, the hosting market  is very heterogeneous, meaning that even shared hosting services can be offered in different variations~\cite{nomshosting2016}. This essentially means that different providers give different administrative rights to their customers (i.e.,  webmasters). 
Second, even if in principle, shared hosting providers leave certain options open to be modified by webmasters, due to the power of default settings, several customers never change those options, even if they can.  
Our manual analysis shows that even if providers do not directly set up a security feature, they can still trigger security measures via  ``recommended settings'' or regularly nudging their customers towards a more secure environment. The same  could hold for  software vendors: we have noticed that for instance, from the latest version onwards, cPanel admin panel removed the server type and version parameter from its default server header.
Third, there is an interaction between some of the features discussed in the Section \ref{sec:features} and content and other applications running alongside a domain, which might require the webmaster to setup certain features such as {\SC} or {\HOC}.

To better capture the role of shared hosting providers in securing their domains while accounting for such interactions, we suggest a different methodology than directly using the features that we have collected.
We  examine  the role of shared hosting providers, by empirically and systematically deducing groups of provider features that correlate strongly together yet vary considerably between providers.
The results of such an approach would then be an empirical recovery of the effects that are throughout the market more dominant, in the realm of shared hosting providers and are either due to the fact that webmasters have no choice or due to default effects, either of which matters significantly.

We do this in two steps: we first use exploratory factor analysis to define latent variables or \emph{factors}. 
Empirically inducing factors from data confirms (or denies) whether the hypothesized division of responsibility is actually present in the population.
We then quantify to what extent each factor is under the control of  shared hosting providers or their customers. 
Note that we purposefully do not use abuse data in this section in order to avoid circular arguments.

\subsection{Exploratory factor analysis}

Factor analysis uses the correlation matrix of all studied variables and reduces its dimensionality  by ``looking for variables that correlate highly with a group of other variables, but correlate very badly with variables outside of that group'' \cite{habing2003exploratory}. The variables with high inter-correlations then shape a factor.
For the factor analysis, we use the security and software features discussed in Section \ref{sec:features}.
Among all our features, the security features  are boolean variables  with 1 pointing to the direction of the variable.
The software features are ordinal from least to most secure with the following order: 0 unpatched versions, 1 patched versions, 2 no software.
Note that in order to simplify the input data, from this section onwards, we consider software with `no version information' as `patched' software with the latest packaged version.
Since our variables are a mix of binary and ordinal, we use Polychoric factor analysis appropriate for ordinal scales.

The input of the factor analysis is an $n \times p$ data matrix with $n$ being the number of measurements (in this case our domains) and $p$ being the number of variables (in this case our features)~\cite{cheung2009malware}.
The factor analysis generates a set of factors, their corresponding factor loadings and factor scores. 
Factor loadings express the relationship (correlation) of each original variable with each factor. Factor scores are the estimated values of the factors per measurement (domain). 
We use parallel analysis for selecting the number of factors, which turns out to be 4.
After applying Varimax factor rotation, we obtain the factor loadings in Table~\ref{tab:factor_loadings}. 
Each row of the table corresponds to a variable, MR1 to MR4 are the factors, and each number indicates the loading of a variable per factor. 
The highest loading per variable is shown in bold. 
Stevens et al. suggest a cut-off point of 0.4 for factor loadings \cite{stevens2012applied}.

\begin{table}[!htp]
\centering
\caption{Output of factor analysis}
\label{tab:factor_loadings}
\resizebox{\columnwidth}{!}{%
\begin{tabular}{lrrrr}
  \toprule
			     & MR1 & MR2 & MR3 & MR4 \\ 
\hline
\hline
\XCTO          &\textbf{0.87} & 0.11  &   0.14 & -0.01 \\
\CSP          &  \textbf{0.80} &0.23 & -0.01 &0.37\\
\XFO           &\textbf{0.83}  & 0.09 &  0.10 & -0.16 \\
\HSTS         &  \textbf{0.61}&  0.50  & 0.04 & 0.03 \\
\MCI          &0.26 &  \textbf{ 0.76} & -0.01 &  -0.24\\
\WBXP         &  -0.39 & \textbf{ 0.68}  &  0.24 & 0.29 \\
\SSVF         & 0.08  &\textbf{ 0.60} & -0.05 & -0.38 \\
\HOC           &  0.13& \textbf{ 0.65} & 0.14 &0.12 \\
\SC            &0.36 & \textbf{0.86}  & 0.03 & 0.11 \\
Patched HTTP*      &0.09  &   0.05 & \textbf{0.74} & -0.11 \\
Secure SSL implementation*      &-0.15 &    -0.09 & \textbf{0.74} & -0.10 \\
Patched SSH*         & -0.07 & 0.04 & \textbf{0.42} & 0.35 \\
Patched PHP*         & 0.09 &-0.12  &  0.13& \textbf{0.55}\\
Patched CMS*         &  -0.14 & 0.01& -0.23& \textbf{0.78} \\
Patched Admin panel*  & 0.08 &  0.08  &0.10 & \textbf{0.58} \\

\midrule
Loadings' sum of squares     &2.90 &2.92 &1.48 &1.90\\
Proportion of variance explained   &0.19 &0.19 &0.10 &0.13\\
Cumulative variance explained   &0.19 &0.39 &0.49 &0.62 \\
\hline
\hline
\multicolumn{5}{l}{* Scale from least to most secure: 0 unpatched, 1 patched or no version,}\\
\multicolumn{5}{l}{~~2 no software}
\end{tabular}
}
\end{table}

The results in Table \ref{tab:factor_loadings} indicate all of the 15 features have a medium to high correlation with  corresponding factors and hence play a significant role in shaping the factors. Factors MR1 to MR4 each explain a part of the total variance. 
The cumulative variance explained in Table \ref{tab:factor_loadings} shows that the four factors together are able to explain 62\% of the variance observed in all the 15 features. This further confirms our earlier call for having four factors, as the majority of variance is captured by them.

From the results it is clear that these four factors (latent variables) capture different aspects of web security. In other words, the factor analysis not only reduce the complexity of our data, but also control for unobserved third variables, as most of the the collected security features do not directly cause abuse. In the following sections we further use these factors to (a) study the respective role of providers and webmasters and 
(b) assess their impact on abuse.


\subsection{Role of Providers in Securing Domains}
\label{subsec:factor_reg}

The combination of the features per factor and their relative loadings (i.e. how much they correlate with different factors)  in Table~\ref{tab:factor_loadings} 
suggest that each of the factors capture a different set of web security efforts.
MR1 consist of features that are partially capturing \textbf{content security practices}. Features in the MR2 factor seem to capture more \textbf{webmaster security practices}. 
Given the high loadings on variables such as {\oHTTP} and {\iSSL}, MR3 clearly captures more \textbf{infrastructure security practices} whereas MR4 seems to capture \textbf{web application security practices}.
In other words, the factor analysis shows that features which one might assume to be related, such as CMS and admin panel, do indeed covary with each other in practice, as they correlate with the same underlying factor.

This leads us to the following hypothesis:
we expect MR1 and MR2 to be be less affected by providers' security efforts than MR3 and MR4.
We examine the relation between the factors and the effort of providers by constructing four linear models.

\begin{table}[!htbp] \centering 
  \caption{Linear Regression Model} 
  \label{tab:reg_factorproviders} 
    \resizebox{\columnwidth}{!}{%
\begin{tabular}{@{\extracolsep{5pt}}lcccc} 
\\[-1.8ex]\hline 
\hline \\[-1.8ex] 
 & \multicolumn{4}{c}{Response Variable: Security Factor(s) } \\ 
\cline{2-5} 
\\[-1.8ex] & MR1 & MR2 & MR3 & MR4 \\ 
\\[-1.8ex] & (1) & (2) & (3) & (4)\\ 
\hline \\[-1.8ex] 
  Hosting provider fixed effect  &yes &yes&yes & yes\\
    & & & & \\ 
    Constant & $-$0.250$^{***}$ & $-$0.300$^{***}$ & 0.100$^{*}$ & 0.420$^{***}$ \\ 
  & (0.064) & (0.066) & (0.043) & (0.051) \\ 
  & & & & \\ 
\hline \\[-1.8ex] 
Observations & 442,075 & 442,075 & 442,075 & 442,075 \\ 
R$^{2}$ & 0.077 & 0.066 & 0.270 & 0.200 \\  
Adjusted R$^{2}$ & 0.075 & 0.064 & 0.270 & 0.200 \\ 
Residual Std. Error (df = 440801) & 1.400 & 1.400 & 0.920 & 1.100 \\  
\hline 
\hline \\[-1.8ex] 
\textit{Note:}  & \multicolumn{4}{r}{$^{*}$p$<$0.05; $^{**}$p$<$0.01; $^{***}$p$<$0.001} \\ 
 & \multicolumn{4}{r}{Standard errors in brackets} \\ 
\end{tabular} }
\end{table} 

To construct these models, we first calculate the factor scores (the estimated values of the factors) from the factor analysis, in a way that a score is assigned to each data point (domain).
We then construct a linear regression model per factor, with the factor score as the dependent variable and provider fixed effects as the independent variable.
The provider fixed effect consists of fitting a separate dummy variable as a predictor for each of the hosting providers in our sample. 
We are interested to see how much of the variance in each of the factors (dependent variables) can be explained by provider efforts, as opposed to individual webmaster efforts. 
The relative difference between the amount of variance explained by each model  indicates the extent that shared hosting providers 
influence the security indicators associated with these factors.


Table \ref{tab:reg_factorproviders} shows the four models and their $R^2$ and adjusted $R^2$ values. To simplify presentation, we omit the estimated coefficient for each hosting provider.
The findings confirm our hypothesis: hosting provider fixed effects explain at least three times more variance in MR3 and partially in MR4 than MR2 and MR1. 
MR1 and MR2, as we earlier hypothesized, with the lowest amount of explained variance, seem to be more a compound of webmaster level efforts rather than provider level influence.
Disregarding the effect of measurement noise, one should note that the $R^2$ value cannot be expected  to be close to 1 in MR3 and MR4, because there are differences between hosting packages offered by different providers.
Similarly, MR1 and MR2 are above zero because customers with specific requirements self-select their provider.


Using the regression results, we are able to empirically confirm our assumptions regarding the role of hosting providers in influencing each of the latent factors constructed using factor analysis. In the following section, we use these results to examine which of the factors have a higher impact on abuse prevalence and which party, provider or webmaster, can influence it more.

\section {Impact of Security Efforts on Abuse}
\label{sec:model}

Having empirically determined the relationship between provider and website security by constructing latent factor, 
we now compare the incidence of abuse at providers to the factors. 
The objective is to test the extent to which the actions of hosting companies and individual webmasters influence the prevalence of abuse, using malware and phishing sites as case studies.

We define our dependent variable $Y_i$ as the number of blacklisted domains in our abuse datasets for $i=1,\ldots,n$, with $n$ being the total number of hosting providers, where $Y_i$ follows a Quasi-Poisson distribution\footnote{We choose Quasi-Poisson over Poisson due to the  over-dispersion (unequal mean and variance) in our data.}. We construct  separate regression models for phishing and malware.

\begin{table}[!htb] 

\begin{minipage}{0.47\textwidth}
\centering 
\setlength\tabcolsep{4pt}
  \caption{Generalized Linear Regression Model (GLM) for count of phishing domains observed per provider} 
  \label{tab:phishingreg} 
  \resizebox{\columnwidth}{!}{%
\begin{tabular}{@{\extracolsep{-4pt}}lccccccc} 
\\[-1.8ex]\hline 
\hline \\[-1.8ex] 
 & \multicolumn{7}{c}{Response Variable: Count of \textbf{phishing} domains} \\ 
\cline{2-8} 
\\[-1.8ex] & \multicolumn{7}{c}{Quasi-Poisson with Log Link Function} \\ 
\\[-1.8ex] & (1) & (2) & (3) & (4) & (5) & (6) & (7)\\ 
\hline \\[-1.8ex] 
 \# domains on shared hosting &  & 1.500$^{***}$ & 1.400$^{***}$ & 1.400$^{***}$ & 1.500$^{***}$ & 1.800$^{***}$ & 1.800$^{***}$ \\ 
  &  & (0.083) & (0.081) & (0.093) & (0.082) & (0.080) & (0.110) \\ 
  & & & & & & & \\ 
\# IPs on shared hosting &  & 0.690$^{***}$ & 0.780$^{***}$ & 0.750$^{***}$ & 0.700$^{***}$ & 0.620$^{***}$ & 0.660$^{***}$ \\ 
  &  & (0.100) & (0.100) & (0.110) & (0.100) & (0.086) & (0.120) \\ 
  & & & & & & & \\ 
 MR1 &  &  & $-$0.570$^{***}$ &  &  &  & $-$0.570$^{*}$ \\ 
 Content security &  &  & (0.140) &  &  &  & (0.240) \\ 
  & & & & & & & \\ 
 MR2&  &  &  & $-$1.100$^{***}$ &  &  & $-$1.100$^{**}$ \\ 
 Webmaster security &  &  &  & (0.270) &  &  & (0.390) \\ 
  & & & & & & & \\
 MR3 &  &  &  &  & $-$0.360$^{**}$ &  & 0.170 \\
 Web infrastructure security &  &  &  &  & (0.110) &  & (0.150) \\ 
  & & & & & & & \\ 
 MR4 &  &  &  &  &  & $-$1.100$^{***}$ & $-$1.200$^{***}$ \\ 
 Web application security &  &  &  &  &  & (0.110) & (0.160) \\ 
  & & & & & & & \\ 
Constant & 3.300$^{***}$ & $-$5.600$^{***}$ & $-$5.700$^{***}$ & $-$5.500$^{***}$ & $-$5.600$^{***}$ & $-$7.100$^{***}$ & $-$7.100$^{***}$ \\ 
  & (0.250) & (0.270) & (0.270) & (0.300) & (0.270) & (0.320) & (0.440) \\ 
  & & & & & & & \\ 
\hline \\[-1.8ex] 
Observations & 1,259 & 1,259 & 1,259 & 1,259 & 1,259 & 1,259 & 1,259 \\ 
\cline{2-8}
 Log Likelihood &-99,401 &-30,094& -29,152 &-28,160 &-29,516&-26,173 &-24,637\\
Dispersion   &2103    & 90  & 88 & 112  & 91 & 75 &  129 \\
Pseudo $R^2$ & - & 0.71 & 0.72 & 0.73 & 0.71  & 0.75 & 0.76  \\ 
Pseudo $R^2$ with regards to model 2 & -	& - &  0.032 & 0.066 & 0.015 & 0.14 & 0.19  \\ 

\hline 
\hline \\[-1.8ex] 
\textit{Note:}  & \multicolumn{7}{r}{$^{*}$p$<$0.05; $^{**}$p$<$0.01; $^{***}$p$<$0.001} \\ 
\end{tabular}
}

\end{minipage}
\hfill
\begin{minipage}{0.47\textwidth}
\setlength\tabcolsep{4pt}
\centering 
  \caption{Generalized Linear Regression Model (GLM) for count of malware domains observed per provider} 
  \label{tab:regmalware} 
    \resizebox{\columnwidth}{!}{%
\begin{tabular}{@{\extracolsep{-4pt}}lccccccc} 
\\[-1.8ex]\hline 
\hline \\[-1.8ex] 
 & \multicolumn{7}{c}{Response Variable: Count of {\bf malware} domains} \\ 
\cline{2-8} 
\\[-1.8ex] & \multicolumn{7}{c}{Quasi-Poisson with Log Link Function} \\ 
\\[-1.8ex] & (1) & (2) & (3) & (4) & (5) & (6) & (7)\\ 
\hline \\[-1.8ex] 
\#  IPs on shared hosting &  & 1.600$^{***}$ & 1.600$^{***}$ & 1.600$^{***}$ & 1.600$^{***}$ & 1.600$^{***}$ & 1.400$^{***}$ \\ 
  &  & (0.090) & (0.090) & (0.087) & (0.089) & (0.098) & (0.095) \\ 
  & & & & & & & \\ 
\# domains on shared hosting &  & 0.460$^{***}$ & 0.560$^{***}$ & 0.520$^{***}$ & 0.470$^{***}$ & 0.480$^{***}$ & 0.600$^{***}$ \\ 
  &  & (0.110) & (0.110) & (0.110) & (0.110) & (0.110) & (0.110) \\ 
  & & & & & & & \\ 
 MR1  &  &  & $-$0.700$^{***}$ &  &  &  & $-$0.310 \\ 
  Content security&  &  & (0.170) &  &  &  & (0.190) \\ 
  & & & & & & & \\ 
 MR2  &  &  &  & $-$1.300$^{***}$ &  &  & $-$1.300$^{***}$ \\ 
 Webmaster security measures &  &  &  & (0.290) &  &  & (0.300) \\ 
  & & & & & & & \\ 
 MR3  &  &  &  &  & $-$0.380$^{**}$ &  & $-$0.130 \\ 
  Web infrastructure security &  &  &  &  & (0.130) &  & (0.130) \\ 
  & & & & & & & \\ 
 MR4  &  &  &  &  &  & -0.170 & -0.360$^{*}$ \\ 
 Web application security &  &  &  &  &  & (0.140) & (0.140) \\ 
  & & & & & & & \\ 
 Constant & 4.300$^{***}$ & $-$4.800$^{***}$ & $-$4.900$^{***}$ & $-$4.600$^{***}$ & $-$4.700$^{***}$ & $-$4.600$^{***}$ & $-$4.300$^{***}$ \\
 & (0.240) & (0.310) & (0.310) & (0.290) & (0.300) & (0.330) & (0.300) \\ 
  & & & & & & & \\ 
\hline \\[-1.8ex] 
Observations & 1,259 & 1,259 & 1,259 & 1,259 & 1,259 & 1,259 & 1,259 \\ 
\cline{2-8}
 Log Likelihood &-273,893 &-79,646  &-75,986  &-73,496  &-78,181  &-79,392  &-71,461\\
Dispersion   &5800    & 330  & 334  & 298 & 332 & 320 &   288 \\
Pseudo $R^2$ & - &  0.71 & 0.73 & 0.74 & 0.72  &  0.71 &  0.74  \\ 
Pseudo $R^2$ with regards to model2 & -	& - &  0.044 & 0.077 &  0.017 & 0.001 & 0.098\\ 
\hline 
\hline \\[-1.8ex] 
\textit{Note:}  & \multicolumn{7}{r}{$^{*}$p$<$0.05; $^{**}$p$<$0.01; $^{***}$p$<$0.001} \\ 
 & \multicolumn{7}{r}{Standard errors in brackets} \\ 
\end{tabular} }

\end{minipage}
\end{table} 

The regression results for phishing and malware abuse are shown in Tables~\ref{tab:phishingreg} and \ref{tab:regmalware}, respectively.
In order to be able to observe the effect of all variables on abuse, we construct one model per variable (models 3-6), together with a final model that includes all variables (model 7). 
We report the dispersion parameter for each of the models. Note that the Quasi-Poisson models are estimated using a Quasi Maximum Likelihood and are adjusted via the reported estimated dispersion peremeter.
Therefore,  the Log Likelihood values are reported from the original Poisson fitted models, as recommended in practice~\cite{quasiposson}.

Moreover, since previous research already established the strong relationship between provider size and abuse prevalence~\cite{toit,armanweis2017},
we use model~2 with only size variables as our base model, and study the extent to which our four factors further explain the variance in abuse, on top of the $R^2$=0.71 of model~(2).
Hence, in addition to the normal pseudo~ $R^2$ value used as a goodness of fit measure for the Quasi-Poisson models~\cite{heinzl2003pseudo}, we report the pseudo~$R^2$ value with respect to model~(2) for each table.

We include both phishing and malware data because while we
see 
some similarities in how abuse type relates to the security characteristics, we also anticipate that there will be differences. Given the specialization in cybercriminal activity, the actors themselves and their preferred methods of compromise are likely different, as is the effectiveness of different security efforts on the side of defenders~\cite{armanweis2017}.

For phishing, three out of four factors are statistically significant when included together in model~(7). Webmaster security (MR2) and  web application security (MR4) play a  statistically significant role  in reducing phishing abuse: for each one unit increase in each of these factors, keeping all other factors constant, phishing abuse drops by $e^{1.100}$ = 3 and $e^{1.200}$ = 3.32, respectively. 

The most prevalent individual indicator that makes up MR2 is the presence of an \texttt{HTTPOnly} cookie, which is a standard XSS defense.
To reiterate, we interpret these features as indicators of a latent factor measuring security effort (see Section~\ref{sec:dead}). For example, the results suggest that when individual webmasters harden the cookie properties of their websites, this is an indication that they also take other (unobservable) measures to inhibit abuse.
The results form MR4 indicates that running up to date versions of CMS and admin panel, or hiding the version information, or running no software, is negatively associated with compromise.
We suspect this is due to the fact that certain providers administer CMSes themselves, to make themselves and their customers less prone to compromise, 
given the vulnerabilities imposed by CMSes and admin panels~\cite{IBMcms,adminpanel}. It also shows that these are the areas that providers' effort can be very effective.

For malware, only MR2 (webmaster security)
and MR4 (Web application security) are significant in model~(7). 
From the two,  webmaster security (MR2) explains most variance in the malware abuse, both when modeled alone (model (4)), and when modeled with other factors (model (7)).
Again, given that \texttt{HTTPOnly} cookie and \texttt{Secure} cookie dominantly shape webmaster security factor (MR2), their significant relation with reducing malware abuse is therefore very intuitive. 
MR4 plays a less significant role in explaining malware abuse. We suspect this is due to the differences in the nature of phishing and malware attacks, attack techniques, and exploited resources. 

Moreover, MR1 (Content security) and MR3 (Web infrastructure security)  show a statistical significant relation with malware and phishing abuse only when considered alone (model (3) and (5), respectively). 
By inspecting regressions including other combinations of factors (not included for space considerations), it appears that MR1 is the more robust indicator than MR3 for the malware regression.

Overall, the combined model explains 19\% of the variance for phishing prevalence and 10\% for malware prevalence among provid-\\ers, beyond the baseline of 71\% explained variance, 
showing that both webmaster and provider efforts influence abuse prevalence.
The influence of these efforts on abuse rates,  for disparate types of abuse (in our case web-based malware and phishing), is consistent in direction
and somewhat varying in magnitude. Finally, we note that while we have explained some of the variation in abuse prevalence among shared hosting providers, much remains unexplained. This should in turn motivate the collection of additional discriminating features in follow-up studies.

\begin{figure}[!ht]
\centering
\subfigure[MR1]
{\includegraphics[width=0.48\columnwidth]{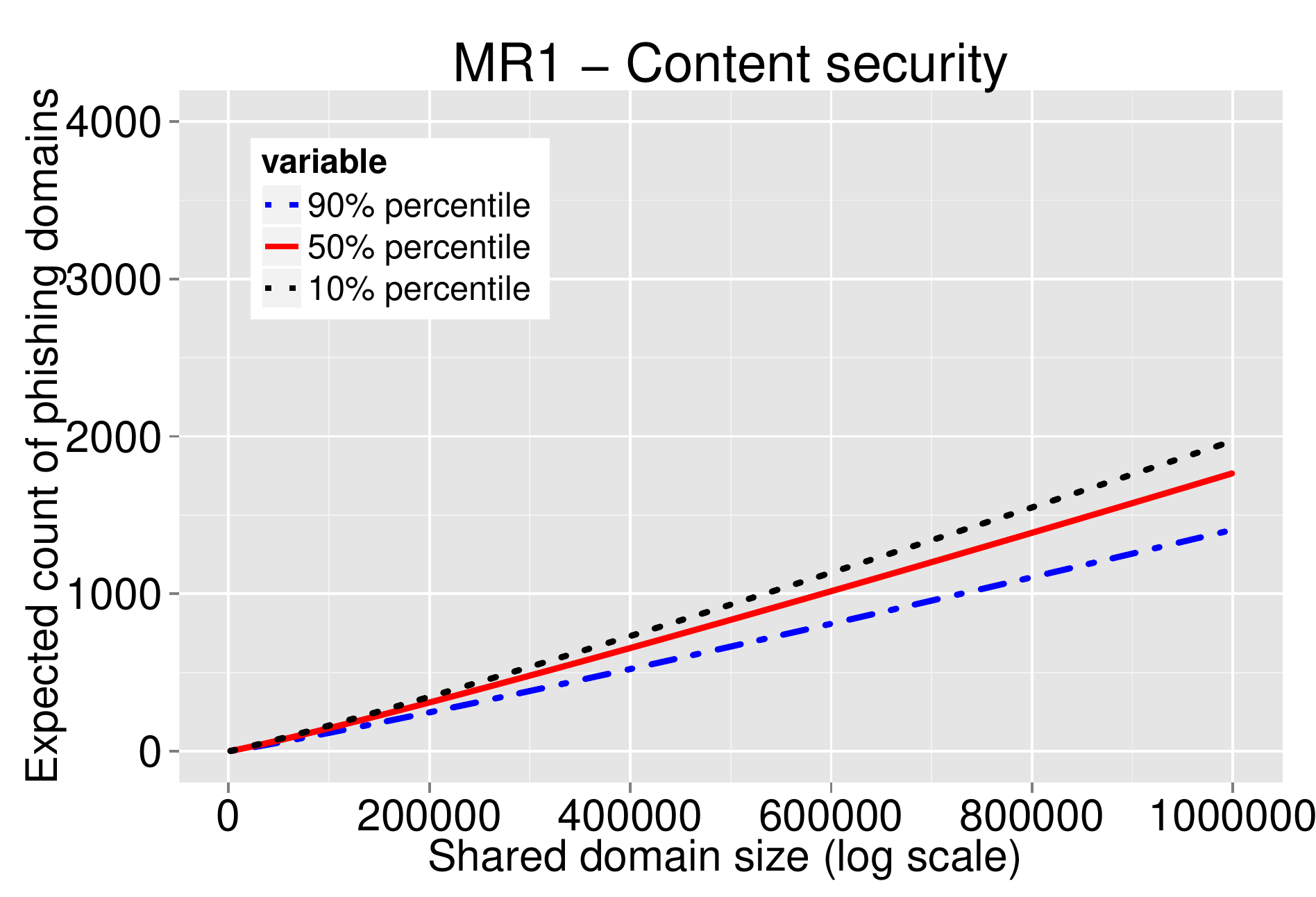}\hfill}
\subfigure[MR2]{ \includegraphics[width=0.48\columnwidth]{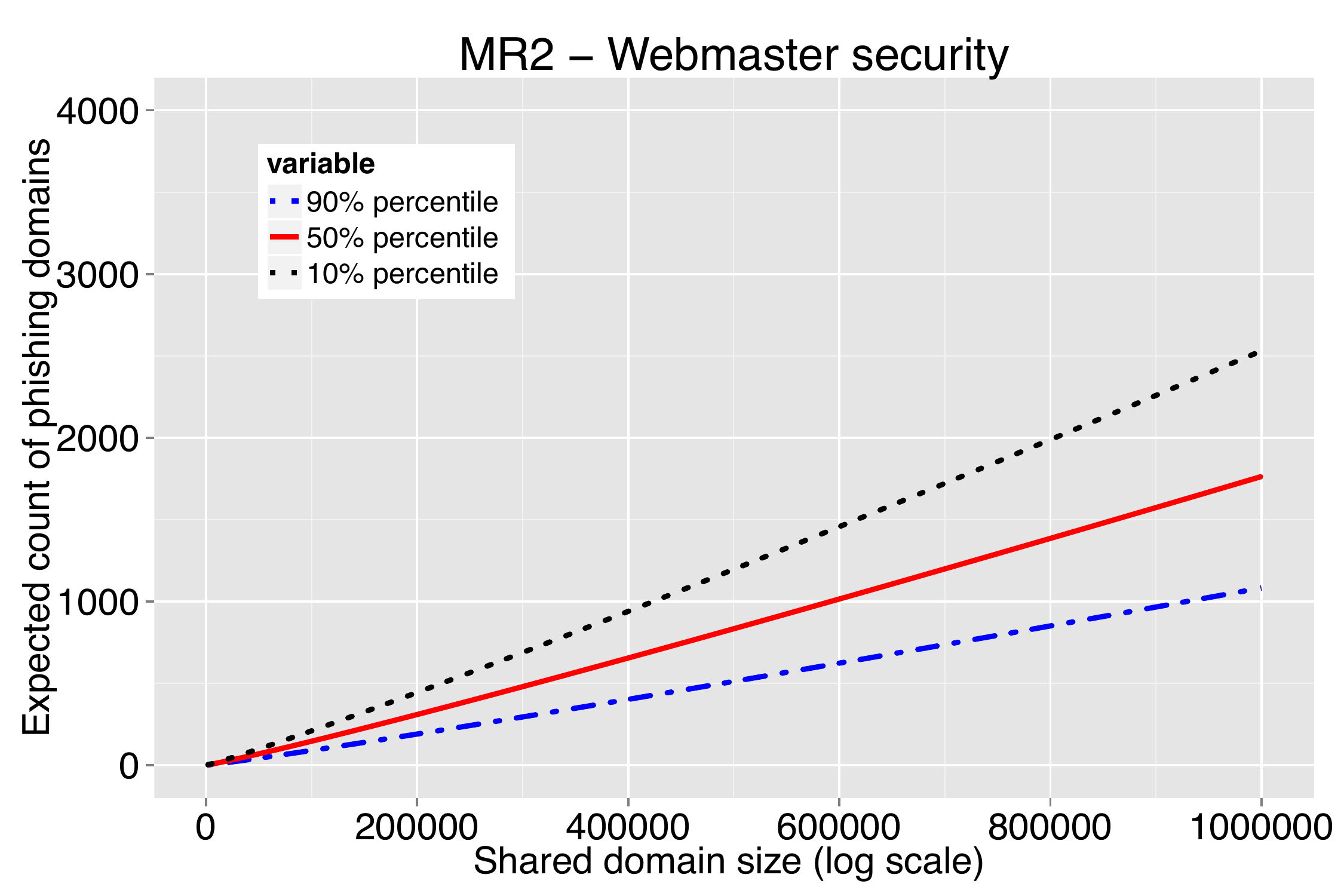}\hfill}
\subfigure[MR4]
{\includegraphics[width=0.48\columnwidth]{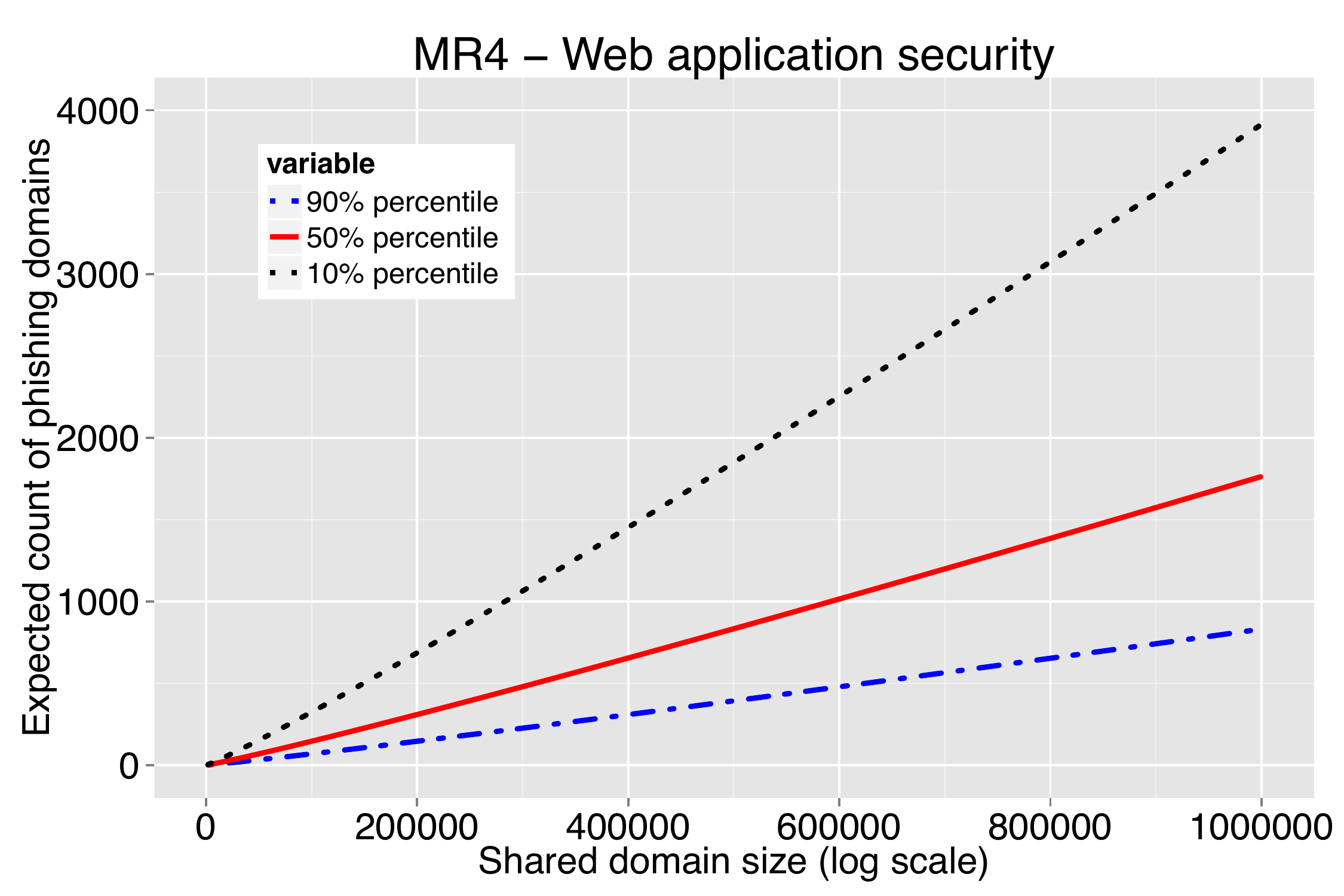}\centering\hfill}
\caption{Plot of expected phishing abuse counts against shared domain size for  MR1, MR2 and MR4 (from model (7) of Table~\ref{tab:phishingreg})}
\label{fig:scenarios}
\end{figure}

Figure~\ref{fig:scenarios} uses the model to demonstrate how the factors influence abuse prevalence. Figure~\ref{fig:scenarios}~(a) plots the expected number of phishing incidents as a function of provider size while varying the value of MR1 (content security) and holding other factors at their median value. Note that we plotted one figure for each of the factors that showed a significant relation with phishing abuse in model~(7) of table~\ref{tab:phishingreg}.
We can see that the bottom 10\% of providers (with the least effort as measured by MR1) should experience less than one and half as many phishing attacks as the top 10\%. In the case of MR2 (webmaster security), the bottom 10\% of providers experience more than twice as many phishing attacks as the top 10\%. For MR4 (web-application security), the difference is even more pronounced: the best-performing 10\% providers by this measure should experience more than 4 times less phishing than the bottom 10\% of providers. 
These findings provide reliable empirical evidence regarding the security benefits of providers adopting industry best practices, most notably proactive patching practices. Given that patching is costly, such evidence is critical to move the industry in the right direction.

\section{Limitations}
\label{sec:limitations}

As with all large-scale studies of real-world applications and implementations, we should reflect on the potential impact of measurement errors and other limitations.
Potential errors in our measurements are caused by the fact that we scan only for the main software packages across the web stack. Also, the collected data can be misinterpreted. One Dutch provider, for example, rolls out its own back-ported security patches for CMSes, without updating the version number. 
Another limitation stems from the use of a rather crude metric for patching (patched/unpatched). An alternative would be looking, for example, at the distance in time between the installed version and the patched versions. 

We captured information on 15 different features, associated with security and patching practices.
Some of these features were very biased, mostly because of their extremely low prevalence. Ideally, we would include features with more variance across the population. 
The features were not interpreted as direct defenses against web compromise, but rather as latent factors that signals effort. However, these feature might also reflect other latent factors in addition to security effort, such as website functionality, popularity, complexity and exposure.

Finally, the reader should bear in mind that our study 
aggregates abuse at the provider level, while features are collected on a separate sample of uncompromised domains in order to increase sample size. Future work could collect features on compromised websites directly to establish stronger differentiation between individual and provider efforts.

\section{Related Work}
\label{sec:related_work}

Because our paper seeks to measure web security in shared hosting environments, identify the role of the hosting providers and its impact on abuse, we build upon several aspects of the literature.

\paragraph{Measuring vulnerabilities of websites and webservers}
There are numerous measurement studies aiming to detect web vulnerabilities across domains (e.g.~\cite{acker,aviram2016drown,Pan,Alhuzali,kals2006secubat,lekies201325,Cangialosi,van2014large,weichselbaum2016csp,calzavara2016content}). For example,
Van Goethem et al.\ assessed 22,000 websites and studied the severity of certain
vulnerability and security features~\cite{van2014large}.
SecuBat developed by Kals et al.\ automatically detects XSS and SQL injection vulnerabilities~\cite{kals2006secubat}.
Lekies et al.\ analyzed the 5,000 most popular domains 
and found that 9.6\% of the examined websites carry at least one DOM-based XSS problem~\cite{lekies201325}. 
Weichselbaum et al.\ detected domains adopting CSP and studied how effective the policies were in protecting against XSS attacks~\cite{weichselbaum2016csp}.
Calzavara et al.\ also studied CSP adoption via a large scale measurement study and concluded that in addition to limited deployment, existing policies are frequently misconfigured~\cite{calzavara2016content}.
Van Acker et al.\ performed a systematic examination of login-page security and found that many login pages are vulnerable to leaking passwords to third parties and to eavesdropping attacks.
They also observed that a few login pages deploy advanced security measures~\cite{acker}.
Lastly, Aviram et al.\ introduced two different attack techniques against SSLv2 and concluded that SSLv2 weaknesses are a significant threat against SSL ecosystem~\cite{aviram2016drown}.



\paragraph{Threats against shared hosting} In addition to general domain vulnerabilities, there are certain threats specific to domains hosted on a shared server. In shared hosting, a physical server is shared among multiple clients, ranging from a few to over a thousand. 
Customers are allocated a fraction of a machine's overall resources and given limited user privileges. Server-side software must be managed by the provider. 
Canali et al.\ examined security performance of a small group of shared hosting providers and concluded that the majority were unable to detect even basic attacks on their infrastructure~\cite{canali2013role}.
The Anti-Phishing Working Group reported that some attackers would compromise shared hosting servers and load phishing pages on each of the hosted websites~\cite{apwg2014}.
Tajalizadehkhoob et al.\ investigated the security performance of different hosting provider types in terms of phishing abuse take-down times and concluded that phishing domains in shared hosting providers often last longer than other group of providers~\cite{nomshosting2016}.
The potential for compromise on a shared environment abuse was first pointed out by Nikiforakis et al.~\cite{nikiforakis2011abusing} and Mirheidari et al.~\cite{mirheidari2012two}, who noted that the lack of enforced session isolation leaves shared web hosts open to mass compromise. Perhaps reflecting this strategy, Vasek et al.\ found that phishing websites were disproportionately likely to be hosted in a shared environment~\cite{vasekriskfactor}.

\paragraph{Relationship between vulnerabilities and abuse} A few studies empirically investigated the relationship between the vulnerabilities of a domain and the likelihood of being compromised.
Vasek and Moore found that Apache and nginx server software and popular CMS platforms, most notably WordPress, Joomla!\ and Drupal, are positive risk factors for webserver compromise~\cite{vasekriskfactor}. 
In fact, a key counterintuitive finding was that fully patched installations have a higher likelihood of compromise than unpatched ones.
Soska and Christin developed an approach that predicts whether websites will be compromised in the near future. The prediction is done via a classifier that is trained on features that are extracted from a set of both malicious and benign websites.
They found CMS type and version to be predictive features, suggesting that many websites could be compromised through a vulnerability in their CMS~\cite{soska2014automatically}.
\paragraph{Role of intermediary in dealing with abuse} 
A number of studies focused on different types of intervention done by intermediaries (e.g.~\cite{Caballero2,moore2007examining,Liu:2011:ERI:1972441.1972448,canali2013role,179521,Cetin2015UnderstandingTR,197118,li2016you,184411,Durumeric:2014}). Moore and Clayton, for example,  
examined the effectiveness of phishing websites take-down by web hosting providers and concluded that website removal is not yet fast enough to completely mitigate the phishing problem~\cite{moore2007examining}. 
Stock et al.\ preformed a large-scale notification campaign of website owners using a set of over 44,000 vulnerable websites and concluded that there are no reliable notification channels that would significantly inhibited the success of notifications~\cite{197118}.
Li et al.\ examined the life cycle of 760,935 hijacking incidents identified by Google Safe Browsing and Search Quality, and found that direct communication with webmasters increased the cleanup rate by 51\%.
They concluded that in order to decrease the number of security incidents, one could increase the webmaster coverage of notification while also equipping hosting providers with tools alerting webmasters to update software~\cite{Li16}.

We build on the existing work in several ways. 
First, we extend the measurement approach developed by Van Goethem et al.~\cite{van2014large} to collect a broader set of features. 
Next,  we move the level of analysis from individual domains to providers. 
In areas beyond shared hosting, researchers have repeatedly found that the intermediaries can play a key role in improving 
security~\cite{van2010role,Liu:2011:ERI:1972441.1972448,McCoy:2012:PRP:2382196.2382285,ClaytonMC15weis,VasekWM16wiscs,armanCSET2015,toit,Tajalizadeh2017Asiaccs}.

Tajalizadehkhoob et al. studied the different factors at work in the abuse data generation process of hosting providers. They identified structural properties and security efforts  of hosting providers, behavior of attackers, and measurement errors,  as factors that can influence concentrations of abuse.
Further, they showed that the structural properties of hosting providers alone -- such as different size, price, and business model variables -- can explain more than 84\% of the variance in abuse concentration of hosting providers~\cite{toit}. 
Noroozian et al. investigated the closely related question of how provider security practices impact abuse concentration and whether the outcome of provider security practices can be indirectly inferred (as a latent variable) from multiple sources of abuse data employing Item-Response Theory~\cite{armanweis2017}. 
Their results quantified the impact of security practices (without knowledge of what those practices may be), demonstrating predictive and explanatory power.
Finally, Sarabi et al. studied the implications of end-user behavior in applying patches. They observed that although both end-users' patching speed and vendors' facilitating policies help in improving the overall  security posture of a host, they were also overshadowed by other factors, such as frequency of vulnerability disclosures and the vendors’ speed in deploying patches~\cite{sarabi2017patch}.

In our study, the hosting company's role is critical, since many domain owners will not be willing or able to adequately secure their site. Our data collection is not based on a random sample from all domains, but on a sampling strategy that covers all shared hosting providers. We present a new approach to disentangle the role of providers and customers in protecting domains. This also allows us to extend the work on the relationship between vulnerability and compromise from the level of individual webmasters to that of providers. 
Last, but not least, we provide the first estimate of the potential gains of such efforts for lowering compromise levels.

\section{Conclusion}
\label{sec:conclusions}

We have undertaken an extensive study of web security efforts.
The purpose of this work is (i) to study the state and landscape of security hygiene at the level of domains and shared hosting providers, (ii) to disentangle the defensive efforts of providers and their customers, and (iii) to assess their impact on web compromise.

Our descriptive findings regarding the web-security landscape shows that most domain security features occur sparsely across the domain and provider space. Even here, though, we see the potential influence of providers. A tiny fraction of providers has very high adoption rates of certain features like {\CSP} and {\HOC}. They appear to offer more managed forms of shared hosting, which might enable them to exert more control over feature configurations of their customers.

Regarding software patching, higher levels of the web stack such as CMS and admin panels are updated more than infrastructure software like SSH and PHP. This might reflect the fact that CMSes and admin panel are attacked more aggressively.
Interestingly, even though infrastructure software is typically under the control of the provider, we see a lot of heterogeneity of versions within the same provider. We suspects this is due to changes in provisioning processes over time. Since patching is costly, earlier default configurations might not get updated unless there is an urgent need.

The individual features should not be interpreted as being directly causing web compromise, for reasons that we laid out in Section \ref{sec:dead}. It is more valid and informative to interpret them as indicators of a latent factor that is the actual causal driver, namely security effort.
Using exploratory factor analysis, we uncovered four such latent factors: content security practices, webmaster security practices, web application security practices and infrastructure security practices. 
The fixed-effect regression analysis uncovered that providers have control over infrastructure and  application security, as we expected. Regarding CMSes specifically, however, the influence of providers is more surprising. This software can run client-side, but still providers influence patch levels. This might mean that a subset of providers administer these installations themselves, or that they  found ways of getting their customers to patch in a timely fashion.

Finally, we model the impact of the four security factors on the compromise rate of providers, as observed in phishing and malware incidents, using Quasi-Poisson GLM regression. Taken together, the  results suggest that both webmaster and provider efforts influence abuse prevalence. While provider security efforts play a more significant role in fighting phishing abuse, webmasters are also effective in reducing abuse rates.
Most of the four factors play a statistically significant role in reducing abuse, either when modeled alone or with other factors.
More specifically, the factor that captures web-master security efforts such as \texttt{Secure} and \texttt{HTTPOnly} cookies, shows a negative relation with both malware and phishing abuse, highlighting the effectiveness of webmasters' efforts in fighting abuse.
The regression results have also shown that web-application security, a factor associated with provider efforts, has a strong significant negative relation with malware and phishing abuse. To illustrate the relative impact, we show that the best-performing 10\% of providers by this measure experience 4 times fewer phishing incidents than the bottom 10\% providers.

In short, our study show that providers have influence over patch levels--even higher in the stack, where CMSes can run as client-side software--and that this influence is tied to a substantial reduction in abuse levels. 
Our study has provided the first rigorous evidence of the security benefits of provider efforts to increase patching levels. This is a critical finding for the dialogue, with and within the industry community, about the merits, costs and benefits of the proposed best practices--e.g., \cite{M3AAWG}.
The takeaway for providers is that improving patch levels pays off. They can do this by administering themselves more of the software installations across the web stack, by securely provisioning default installations or by deploying some other mechanisms that enable them to get their customers to collectively reach higher patch levels.

Beyond the area of shared hosting and web compromise, our study provides a new methodological approach to disentangle the impact of different actors on security. This approach can be adopted to study other areas of joint responsibility, such as between cloud hosting providers and tenants, or corporate system administrators and end users.

Measuring effort in a heterogeneous environment with different requirements is hard.
Future work could measure feature use before (or together with) security. Measuring security alone is vulnerable to spurious correlations and inferences, when not controlling for the differences in website functionality, complexity, exposure, et cetera. Another future direction is to make this approach longitudinal, in order to tell apart which fraction of security effort is reactive (i.e., reacting to compromise) and to better detect the direction of causality.
In the end, we hope to provide better empirical support for industry best practices focused on hosting providers.

\begin{acks}
The authors thank Farsight Security for providing access to DNSDB data.  This work was supported by NWO (grant nr. 12.003/628.001.003), the Dutch National Cyber Security Center (NCSC) and SIDN, the \texttt{.NL} Registry, and Archimedes Privatstiftung, Innsbruck.
Additionally, we thank  our ACM CCS reviewers for their  useful feedback and support in improving the paper for the camera-ready version.
\end{acks}


\bibliographystyle{ACM-Reference-Format}
\bibliography{Bibliography}


\appendix
\section{Version Information Details }
\label{app:ver}
\begin{table}[!htb] 
	\centering 
	\caption{The list of versions per software that are considered patched (patched = latest packaged version in Ubuntu, Debian or CentOS)} 
	\label{} 
	  \resizebox{\columnwidth}{!}{%
	\begin{tabular}{@{\extracolsep{5pt}}lr} 
		\\[-1.8ex]\hline 
		\hline \\[-1.8ex]
		Software & Version considered patched \\
		\toprule
		Apache    & [2.2.15 - 2.2.22 -  2.4.7 - 2.4.10 - 2.4.18 - 2.4.20 - 2.4.23] \\
		SSH & [5.3p1 - 5.9p1 - 6.0p1 - 6.6p1 - 6.6.1p1 - 6.7p1 - 7.1p2 - 7.2p2 - 7.3 - 7.3p1] \\
		WordPress & [4.7 - 4.6.1 - 4.5.4 - 4.4.5 - 4.3.6
4.2.10 - 4.1.13 - 4.0.13
3.9.14 - 3.8.16 - 3.7.16]\\
		Joomla! & [3.6.4]\\
		Drupal & [7.52 - 8.2.3]\\
		cPanel & [7.52]\\
		DirectAdmin & [1.50.1]\\
		Virtualmin & [1.820]\\
		Plesk & [12.5.30 - 17.0.16]\\
		Microsoft IIS & [12 - 10 - 9 - 8.5]\\
		Nginx & [1.2.1 - 1.4.6 - 1.10.0 - 1.10.1 - 1.10.3 - 1.11.5] \\
		PHP & [5.3.10 - 5.3.3 - 5.4.45 - 5.5.9 - 5.6.27 - 5.6.28 - 6.6.30 - 7.0.11 - 7.0.12 -7.0.13]\\
	\end{tabular}}
\end{table}

\end{document}